\documentclass[%
reprint,
superscriptaddress,
amsmath,amssymb,
aps,
floatfix,
]{revtex4-1}

\usepackage{graphicx}% Include figure files
\usepackage{subcaption}
\usepackage{rotating}
\usepackage{dcolumn}% Align table columns on decimal point
\usepackage{bm}% bold math
\usepackage{braket}
\usepackage{amsmath}
\usepackage{mathtools}
\usepackage[colorlinks,allcolors=blue]{hyperref} 
\usepackage{hyperref}% add hypertext capabilities
\usepackage{natbib}
\begin{document}
	
	\raggedbottom
	
	\preprint{APS/123-QED}
	
	\title{Ultra-High-Precision Detection of Single Microwave Photons based on a Hybrid System between Majorana Zero Mode and a Quantum Dot}
	
	\author{Eric Chatterjee}
	\affiliation{Sandia National Laboratories, Livermore, California 94550, USA}
    \author{Wei Pan}
	\affiliation{Sandia National Laboratories, Livermore, California 94550, USA}
	\author{Daniel Soh}
	\affiliation{Sandia National Laboratories, Livermore, California 94550, USA}

	\date{\today}% It is always \today, today,
	
	\begin{abstract}
		The ability to detect single photons has become increasingly essential due to the rise of photon-based quantum computing. In this theoretical work, we propose a system consisting of a quantum dot (QD) side-coupled to a superconducting nanowire. The coupling opens a gap in both the QD mode and the Majorana zero mode (MZM) at the nanowire edge, enabling photon absorption in the system. We show that the absorbed photoelectron decays via rapid (sub-nanosecond to nanosecond) nonradiative heat transfer to the nanowire phonon modes rather than by spontaneous emission. Furthermore, we calculate the temperature increase and associated resistance increase induced by the absorption of a photon for a given appropriate set of material and environmental parameters, yielding a temperature increase in the millikelvin range and a resistance increase in the kiloohm range, vastly exceeding the photon-absorption-induced temperature and resistance increases for competing 2D-3D hybrid systems by 5 and 9 orders of magnitude, respectively. Lastly, we determine the detector efficiency and discuss the system density required for deterministic photon number measurement, demonstrating that a photon absorption probability of over 99.9 percent can be achieved for an integrated system consisting of an array of nanowire-QD complexes on-chip inside a cavity. Our results thus provide a basis for a deterministic microwave photon number detector with an unprecedented photon-number-detection resolution.
	\end{abstract}
	
	\pacs{Valid PACS appear here}% PACS, the Physics and Astronomy
	\maketitle

\section{Introduction}     % section 1.1

The ability to detect single radio-frequency photons has become increasingly essential due to the rise of superconducting quantum computing. The bolometer approach of single-photon detectors is more appropriate than others because of the extreme sensitivity of temperature-change-induced physical measurement capabilities \cite{chatterjee2021microwavephoton, irwin1995applicationelectrothermal, miller2003photoncounter, lita2008countingphotons}. The key requirements in designing a bolometer-type photon number detector are to ensure spatial separation between the absorber (the part of the system absorbing the photons) and the bolometer (the part that is used as a platform for measurement) so that the measurement process does not wash out the absorbed photoelectron, rapid and deterministic energy transfer from the absorber to the bolometer so that the parasitic radiative decay process does not annihilate the excited photoelectron before the measurement can be performed, and high-precision resolution, given the relatively small energy of a single radio-frequency photon. Past proposals have focused on hybrid 2D/3D systems, with a 2D surface state absorbing the ambient photons and subsequently transferring energy through heat transfer to the 3D bulk phonon modes, upon which the energy gain in the bulk is measured via bolometry. Examples of this setup include transition edge sensors (TESs) \cite{irwin1995applicationelectrothermal, miller2003photoncounter, lita2008countingphotons}, and more recently, Dirac semimetals such as Cd\textsubscript{3}As\textsubscript{2} \cite{chatterjee2021microwavephoton} consisting of a proximity-induced superconducting bulk (with a gap larger than the microwave photon frequency) and a graphene-like topological surface state. Although such systems achieve absorber-bolometer spatial separation, with the Cd\textsubscript{3}As\textsubscript{2} detector achieving rapid energy transfer from the photoelectrons to the bolometer as well, the large volume required for the hybrid 2D/3D systems sharply limits the temperature increase per absorbed photon, with sub-micro-ohm measurement required to resolve the temperature-induced resistance increase in the bulk. 

In order to achieve a measurement resolution sufficient for detecting every absorbed single photon, it is therefore desirable to use a low-dimensional detector that reduces vastly the heat capacity and, thus, increases the temperature contrast. To this end, superconducting nanowires (consisting of a 1D semiconductor that has acquired a superconducting gap through proximity) provide an ideal platform for single photon detection. Superconducting nanowire single photon detectors (SNSPDs), which involve a photon absorbed by a Cooper pair causing the nanowire to revert to the normal (non-superconducting) state, thereby reducing the current flow through the wire, have gained popularity as a means of detecting optical photons \cite{jaspan2006heraldingphoton, hadfield2006quantumkey, hadfield2009singlephoton, nataranjan2012superconductingnanowire, zadeh2021superconductingnanowire}. However, due to efficiency constraints at lower photon frequencies, research thus far has been limited to telecommunication wavelengths \cite{hu2009fibercoupled, miki2010multichannelsnspd}, or more recently infrared wavelengths \cite{marsili2013detectingsingle}. 

Here, we propose a system consisting of a $p$-wave superconducting nanowire side-coupled to a quantum dot at each end. The topological edge state of such a nanowire has been theorized to be a Majorana bound state \cite{sau2010genericplatform, dassarma2015majoranamodes}, which is by itself incapable of absorbing single photons due to the lack of an electric dipole moment. However, the hybrid Majorana-QD mode can absorb single photons, exciting the system to the higher-energy state consisting of a superposition of an excited QD electron and a nanowire edge state excitation. Ideally, the excitation would then decay to the ground state via nonradiative heat transfer to the phonon modes of the nanowire. The consequent temperature increase in the wire can then be determined by measuring the increase in the longitudinal resistivity. We will calculate the resistance increase per photon as a function of the sample dimensions and material properties, with the goal of ensuring high-precision resolution for the detector. In order to measure the longitudinal resistance without perturbing the Majorana modes (thus ensuring absorber-bolometer separation), we will place each lead at least 150 nm inward from the corresponding edge.  Furthermore, we will theoretically derive the nonradiative energy transfer time from the QD-Majorana mode to the nanowire phonons, so that we can compare to the time for the undesired radiative decay process. 

Our ultra-high-precision microwave photon number detector serves as a major breakthrough in multiple respects. It is the first real-world application of Majorana zero modes other than topological quantum computing. It also provides a revolutionary improvement in detection resolution for microwave photons, with a 9-orders-of-magnitude resolution improvement over a Cd\textsubscript{3}As\textsubscript{2} detector \cite{chatterjee2021microwavephoton}. Finally, the system is highly integrable due to the extremely small size of each QD-Majorana complex.

The paper is organized as follows: In Sec.~\ref{sec: QD-Majorana Hybrid States and Photon Absorption}, we derive the QD-Majorana hybridized spectrum and the interaction strength between a photon and an electron in this complex. In Sec.~\ref{sec: Temperature Increase Per Absorbed Photon}, we calculate the heat capacity of the nanowire, and from that the temperature increase per absorbed photon. Section~\ref{sec: Energy Transfer Rate from QD-Majorana to Nanowire Bulk} shows the method for calculating the energy transfer rate from the excited photoelectron to the nanowire phonons. In Sec.~\ref{sec: Ensuring Deterministic Photon Number Detection}, we lay out the means of calculating the photon absorption probability for a system of QD-nanowire complexes on-chip inside a microwave cavity. Finally, in Sec.~\ref{sec: Optimizing Parameters}, we optimize the numerical values for the parameters. The results demonstrate a near-deterministic photon absorption probability of over 99.9\%, an absorber-to-bolometer energy transfer rate exceeding the parasitic radiative decay rate by over 8 orders of magnitude, and most crucially, an over 6-order-of-magnitude improvement in single-photon resolution compared to a 2D/3D hybrid system such as a Cd\textsubscript{3}As\textsubscript{2} detector \cite{chatterjee2021microwavephoton}.

\section{QD-Majorana Hybrid States and Photon Absorption}
\label{sec: QD-Majorana Hybrid States and Photon Absorption}

\begin{widetext}
\begin{figure*}[!tb]
	\centering
	\includegraphics[width=0.8\linewidth]{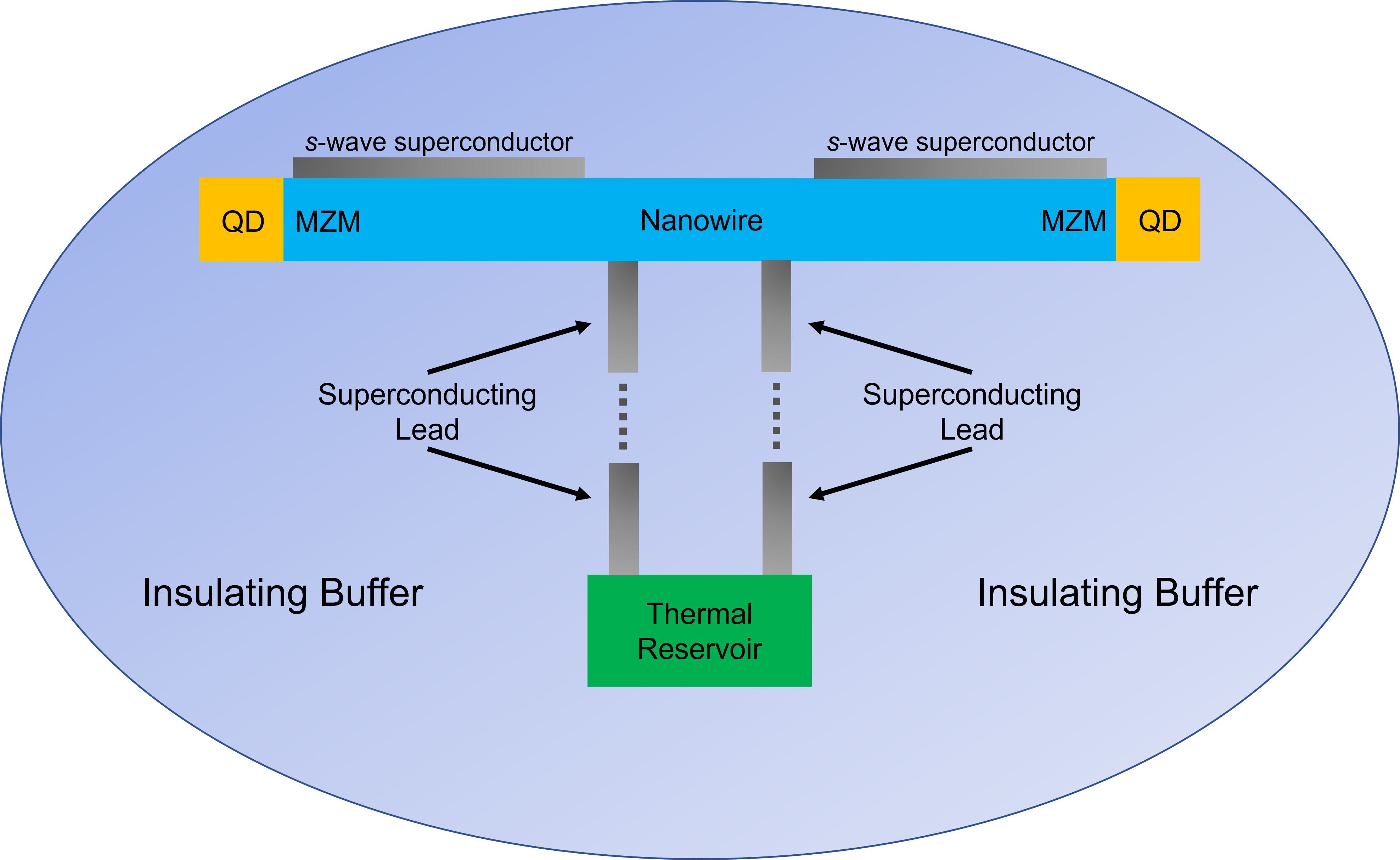}
	\caption{(Wider figure) System diagram. The Majorana zero mode (MZM) at each end of the nanowire is coupled to a quantum dot (QD). Note that the figure is not drawn to scale.}
	\label{fig:systemdiagram}
\end{figure*}
\end{widetext}

The setup of the QD-nanowire system is depicted in Fig.~\ref{fig:systemdiagram}.
At low temperatures, the $s$-wave superconducting strips induce $p$-wave superconductivity in the semiconducting nanowire, leading to the formation of Majorana zero modes (MZMs) at the edges. Fundamentally, this process is driven by two phenomena. First, Rashba spin-orbit coupling, as well as Zeeman splitting induced by an applied magnetic field, give rise to spin-momentum locking for the semiconducting nanowire at the Fermi level. Next, Cooper pairs tunnel from the $s$-wave superconducting strips to the semiconducting nanowire due to the proximity effect. The former and latter effects are represented by the following Hamiltonian operators $H_0$ and $H_\mathrm{SC}$, respectively \cite{sau2010nonabelian, lutchyn2010majoranafermions}:
\begin{align}
H_0 &= \frac{p^2}{2m} - \mu + V_z \sigma_z + \alpha \sigma_y p_x, \\
H_\mathrm{SC} &= \int dx \Big(\Delta(x) f_{\uparrow}^{\dag}(x) f_{\downarrow}^{\dag}(x) + \Delta^*(x) f_{\downarrow}(x) f_{\uparrow}(x)\Big),
\end{align}
where the constants $m$, $\mu$, $V_z$, and $\alpha$ represent the effective electron mass, chemical potential, effective Zeeman splitting, and Rashba spin-orbit interaction strength, respectively, the operators $p$ and $\sigma$ denote the momentum and spin, respectively, and the spatially-varying parameter $\Delta(x)$ is the proximity-induced superconducting gap. Note that $\hat{x}$ is defined as the axis of the nanowire, whereas $\hat{z}$ is the direction of the applied magnetic field (i.e., the axis perpendicular to the interface between the semiconducting nanowire and the $s$-wave strips).

\begin{figure}[!tb]
	\centering
	\includegraphics[width=\linewidth]{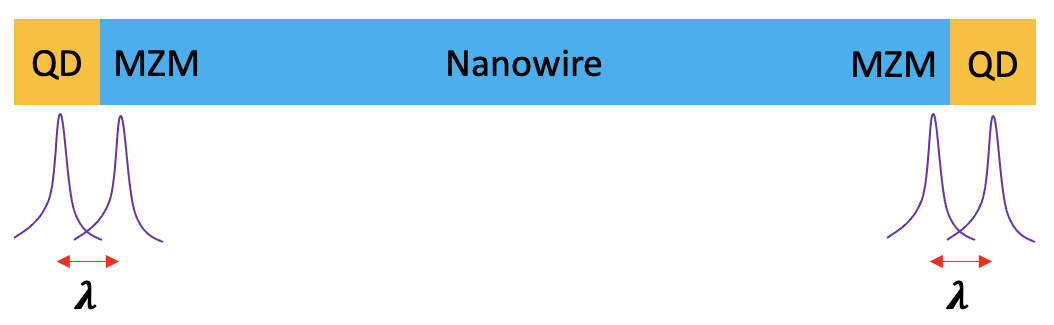}
	\caption{Depiction of the coupling between each quantum dot (QD) and the adjacent Majorana zero mode (MZM), with $\lambda$ representing the hopping parameter between QD and MZM wavefunctions.}
	\label{fig:qdnanowirediagram}
\end{figure}

A pair of leads are connected at the ends of the intermediate region between the $s$-wave strips in order to dynamically measure the resistance increase (which serves as a proxy for temperature increase). The nanowire is side-coupled to a QD at each end, with a hopping parameter of $\lambda$ between a QD mode and the Majorana zero mode (MZM) facing the QD, as depicted in Fig.~\ref{fig:qdnanowirediagram}.
The length of the nanowire is far greater than the coherence length of the Majorana wavefunction, thus rendering any coupling between the opposite-end Majorana modes negligible. For each quantum dot, we start by setting the gate voltage such that the its energy level (when occupied) is aligned with that of the Majorana state. We set the ladder operators for the quantum dot with Fock states $\ket{0}$ (unoccupied) and $\ket{1}$ (occupied) as:
\begin{align}
d = \ket{0}\bra{1}, \\
d^{\dag} = \ket{1}\bra{0}.
\end{align}
Next, we consider the Majorana operators on each end of the nanowire. In general, for an $N$-site chain, the unpaired ``left" and ``right" Majorana operators ($\eta_1^L$ and $\eta_N^R$, respectively) are defined in terms of the fermionic operators $f$ for the first and $N^{\textrm{th}}$ sites as follows \cite{shapourian2017manybody}:
\begin{align}
\label{eq: left Majorana operator}
\eta_1^L &= f_1 + f_1^{\dag}, \\
\label{eq: right Majorana operator}
\eta_N^R &= -i(f_N - f_N^{\dag}).
\end{align}
Note that $\Big(\eta_1^L\Big)^2 = I = \Big(\eta_N^R\Big)^2$, indicating that the Majorana operators cycle the system between states in a twofold-degenerate system (assuming that the nanowire is much longer than the Majorana wavefunction's coherence length). Given the requirement that these zero-energy states obey particle-hole symmetry \cite{leumer2020exacteigenvectors}, the edge modes will be composed of equal-weight superpositions of unoccupied and occupied regular fermionic states. Specifically, it is worth noting that the two Majorana superpositions for each edge feature opposite parities. Applying the particle-hole symmetry condition, the states take the following form \cite{shapourian2017manybody}:
\begin{align}
\ket{\psi_{\pm}}_1 &= \frac{1}{\sqrt{2}} \Big(\ket{0}_1 \pm \ket{1}_1\Big), \\
\ket{\psi_{\pm}}_N &= \frac{1}{\sqrt{2}} \Big(\ket{0}_N \pm \ket{1}_N
\Big),
\end{align}
Intuitively, for each edge, the $\ket{0}$ and $\ket{1}$ states can be conceptualized as a half-hole and a half-electron, respectively. As such, the states satisfy the requirement that a Majorana state be composed of an electron-hole superposition \cite{alicea2012newdirections, beenakker2014annihilationquasiparticles}. As desired, the Majorana operators act to flip the state parities:
\begin{align}
\eta_1^L \ket{\psi_{\pm}}_1 &= \pm \ket{\psi_{\pm}}_1, \\
\eta_N^R \ket{\psi_{\pm}}_N &= \mp i \ket{\psi_{\mp}}_N.
\end{align}
The overlap between the Majorana wavefunction and the quantum dot wavefunction gives rise to the following interaction Hamiltonian \cite{liu2011detectingmajorana} for the left and right edges, respectively:
\begin{align} 
\label{eq: QD-Majorana Hamiltonian left}
H_L &= \hbar \lambda (d + d^{\dag}) \eta_1^L, \\
\label{eq: QD-Majorana Hamiltonian right}
H_R &= -i \hbar \lambda (d - d^{\dag}) \eta_N^R,
\end{align}
where $H_R$ is modified to ensure that the interaction Hamiltonian only features real matrix elements corresponding to the hopping between $\ket{01}$ and $\ket{10}$, i.e., between the QD wavefunction and the Majorana wavefunction, and that these hopping parameters equal those for $H_L$. This in turn ensures that these wavefunctions remain real-valued in the real-space representation, thus simplifying the dipole matrix element calculation that will be performed later in this work.

For the left edge, in the composite basis $\Big(\ket{0\psi_-}_L,\ket{0\psi_+}_L, \ket{1\psi_+}_L, \ket{1\psi_-}_L\Big)$, the QD-MZM interaction Hamiltonian is represented by the following matrix:
\begin{equation}
H_L = \hbar \lambda
\begin{pmatrix}
0 & 0 & 0 & -1 \\
0 & 0 & 1 & 0 \\
0 & 1 & 0 & 0 \\
-1 & 0 & 0 & 0
\end{pmatrix}.
\end{equation}
It is evident that the Hamiltonian couples $\ket{0\psi_{\pm}}_L$ with $\ket{1\psi_{\pm}}_L$, thus allowing the composite 4-dimensional system to be split into two 2-dimensional subsystems. Physically, we can explain it as the QD-Majorana interaction switching the QD state while conserving the Majorana state. Solving for the eigenvalues and eigenvectors of $H$, we find that the energy levels become the following:
\begin{equation}
E_{\pm} = \pm \hbar \lambda,
\end{equation}
corresponding to the following composite states:
\begin{align}
\label{eq: A,+ state}
\ket{A,+} &= \frac{1}{\sqrt{2}} \Big(\ket{0\psi_-}_L - \ket{1\psi_-}_L\Big), \\
\label{eq: B,+ state}
\ket{B,+} &= \frac{1}{\sqrt{2}} \Big(\ket{0\psi_+}_L + \ket{1\psi_+}_L\Big), \\
\label{eq: A,- state}
\ket{A,-} &= \frac{1}{\sqrt{2}} \Big(\ket{0\psi_-}_L + \ket{1\psi_-}_L\Big), \\
\label{eq: B,- state}
\ket{B,-} &= \frac{1}{\sqrt{2}} \Big(\ket{0\psi_+}_L - \ket{1\psi_+}_L\Big),
\end{align}
where each state is labeled $+$ or $-$ based on whether the corresponding energy is $+\hbar \lambda$ or $-\hbar \lambda$.

\begin{figure*}[!tb]
	\centering
	\includegraphics[width=\textwidth]{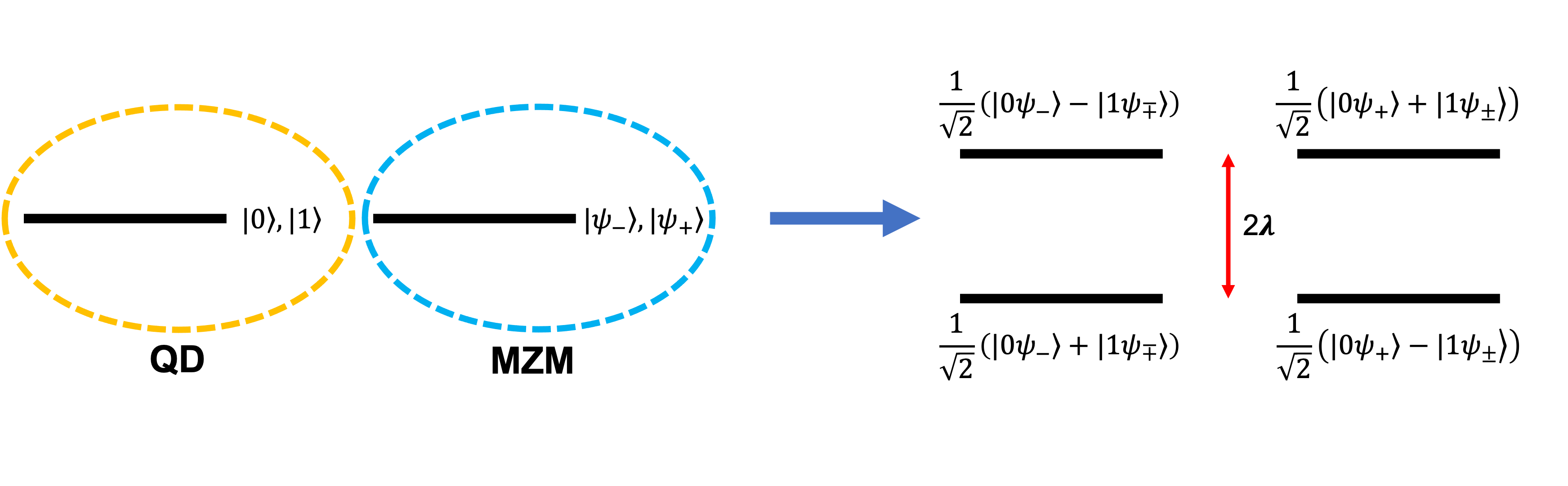}
	\caption{Depiction of hybridization process between the quantum dot (QD) state and the Majorana-zero-mode (MZM) state at the each edge of the nanowire, with the upper (lower) sign for the composite states corresponding to the left (right) edge. Note that an energy gap resonant with a photon frequency of $2\lambda$ is opened, and that the higher-energy and lower-energy states feature opposite phases.}
	\label{fig:qdnanowireenergylevels}
\end{figure*}

For the right edge, in the composite basis $\Big(\ket{0\psi_-}_R,\ket{0\psi_+}_R, \ket{1\psi_-}_R, \ket{1\psi_+}_R\Big)$, the interaction Hamiltonian takes the following matrix form:
\begin{equation}
H_R = \hbar \lambda
\begin{pmatrix}
0 & 0 & 0 & -1 \\
0 & 0 & 1 & 0 \\
0 & 1 & 0 & 0 \\
-1 & 0 & 0 & 0
\end{pmatrix}.
\end{equation}
As with the left edge, the Hamiltonian couples $\ket{0\psi_{\pm}}_R$ with $\ket{1\psi_{\mp}}_R$, allowing the composite 4-dimensional system to be split into two 2-dimensional subsystems. Here, too, we find that the energy levels split into $\pm \hbar \lambda$, corresponding to the following composite states:
\begin{align}
\label{eq: C,+ state}
\ket{C,+} &= \frac{1}{\sqrt{2}} \Big(\ket{0\psi_-}_R - \ket{1\psi_+}_R\Big), \\
\label{eq: D,+ state}
\ket{D,+} &= \frac{1}{\sqrt{2}} \Big(\ket{0\psi_+}_R + \ket{1\psi_-}_R\Big), \\
\label{eq: C,- state}
\ket{C,-} &= \frac{1}{\sqrt{2}} \Big(\ket{0\psi_-}_R + \ket{1\psi_+}_R\Big), \\
\label{eq: D,- state}
\ket{D,-} &= \frac{1}{\sqrt{2}} \Big(\ket{0\psi_+}_R - \ket{1\psi_-}_R\Big).
\end{align}
The hybridization of the states is depicted in Fig.~\ref{fig:qdnanowireenergylevels}.
Conceptually, the energy level shift relative to an uncoupled system can be understood as being distributed across the quantum dot mode and the Majorana bound state, thus corresponding to the splitting of the Majorana zero mode.

We are now ready to consider the photon absorption by the QD-Majorana coupled system, provided a photon field resonant with the gap between the $-$ and $+$ QD-Majorana states, i.e. $\omega = 2\lambda$. The electric field $\bm{E}$ interacts with the dipole moment $\bm{d}$ of the QD-Majorana complex via the following perturbative Hamiltonian:
\begin{equation}
H' = -\bm{d} \cdot \bm{E}.
\end{equation}
The dipole moment calculation is discussed in detail in Appendix~\ref{sec: Calculating the Dipole Matrix Elements}. For the transmon frequency of about 5 GHz, the results reveal a dipole matrix element amplitude of $|d_{+,-}| \approx 1.8 \times 10^{-26} \textrm{ C} \cdot \textrm{m}$ for a QD coherence length matching that of the Majorana wavefunction, corresponding a QD-Majorana center-to-center distance of 440 nm.

On the other hand, the photon field can be expanded in terms of the ladder operators $a^{(\dag)}$ in the conventional manner:
\begin{equation}
\bm{E} = \bm{E_\mathrm{zpf}} \Big(a + a^{\dag}\Big),
\end{equation}
where the electric field zero-point function $\bm{E_\mathrm{zpf}}$ is oriented along the field polarization axis and carries the following amplitude as a function of the frequency $\omega$ and cavity volume $V$ \cite{kakazu1994quantizationelectromagnetic}:
\begin{equation}
E_\mathrm{zpf} = \sqrt{\frac{\hbar \omega}{\epsilon_0 V}}
\end{equation}
The Jaynes-Cummings Hamiltonian $H'$ can thus be expressed in terms of the ladder operators connecting lower-energy and higher-energy QD-Majorana states $\ket{n_i,-}$ and $\ket{n_f,+}$ (respectively) as interactions with coupling coefficient $g_{n_f,n_i}$:
\begin{equation}
H' = \hbar \sum_{n_i,n_f} g_{n_f,n_i} \Big(a^{\dag} b_{n_f,n_i} + a b_{n_f,n_i}^{\dag}\Big),
\end{equation}
where we sum over all possible values of $n_i$ and $n_f$ from the list $(A,B,C,D)$, and the operator $b_{n_f,n_i}$ is defined as follows:
\begin{equation}
b_{n_f,n_i} = \ket{n_i,-} \bra{n_f,+},
\end{equation}
and the interaction coefficients $g_{n_f,n_i}$ feature the following amplitude:
\begin{align} \label{eq: QD-photon coupling coefficient}
\begin{split}
|g| &= \Big|\braket{n_i,-|-\bm{d} \cdot \bm{\hat{E}}|n_f,+}\Big| \frac{E_\mathrm{zpf}}{\hbar} \\
&= |d_{+,-}| \sqrt{\frac{\omega}{\hbar \epsilon_0 V}}.
\end{split}
\end{align}
In the last line, we have assumed that the polarization axis of the electric field is in line with the center-to-center axis between the QD and the Majorana mode, which represents the QD-Majorana dipole axis (effectively, the polarization axis for the QD-Majorana system). Once the system has been photoexcited, we desire for the excitation to rapidly decay via heat transfer to the nanowire phonon modes, while avoiding radiative decay. Quantitatively, this corresponds to a nonradiative decay rate $\Gamma_\mathrm{nr}$ much greater than the radiative decay rate $\Gamma_\mathrm{rad}$, i.e. $\Gamma_\mathrm{nr} \gg \Gamma_\mathrm{rad}$. On the other hand, it is also important to ensure that the nonradiative loss rate is well below the gap frequency, i.e. $\Gamma_\mathrm{nr} \ll \omega$, so that the spectral broadening does not wash out the distinctness of the states. Furthermore, we also wish to ensure that a measurable temperature increase is registered in the nanowire for a single-photon absorption by the QD-Majorana system.

\section{Temperature Increase Per Absorbed Photon}
\label{sec: Temperature Increase Per Absorbed Photon}

Here, we determine the heat capacity of the superconducting nanowire in order to calculate the temperature increase caused by the absorption of a single photon by the QD-Majorana system. In the low-temperature limit, the occupied phonon modes are restricted to the long-wavelength acoustic modes, thus enabling the use of the Debye model \cite{debye1912zurtheorie}. The nanowire contains 1 longitudinal acoustic branch and 2 transverse (torsional) acoustic branches \cite{mante2018acousticphonons}. To first order, these branches feature approximately equal speeds of sound, and we label the average value as $v_s$. Applying the Bose-Einstein distribution, the total phonon energy equals the following summation over the wavevectors $\bm{q}$:
\begin{equation} \label{eq: U_ph summation}
U_{ph}(T) = 3 \sum_{\bm{q}} \frac{\hbar \omega_q}{e^{\frac{\hbar \omega_q}{k_B T}} - 1},
\end{equation}
where $\omega_q = v_s q$ represents the average frequency of a mode at wavevector $\bm{q}$. As in a 3D lattice, this summation can be solved by determining the density of states for each branch, given a 1D lattice of length $L$:
\begin{equation}
D(\omega) = 2 \frac{dN}{dq} \frac{dq}{d\omega} = 2 \frac{L}{2\pi} \frac{1}{v_s} = \frac{L}{\pi v_s}. \label{eq: phonon density of states}
\end{equation}
Note that the factor of 2 in the first line is inserted in order to ensure that for each wavevector amplitude $q$, the modes at both $+\bm{q}$ and $-\bm{q}$ are included. 

We thus convert the summation in Eq.~\eqref{eq: U_ph summation} to integral form by incorporating the density of states. Since only the linear regime of the acoustic branches is non-negligibly occupied at low temperatures, we can integrate to infinite energy without measurable loss of accuracy:
\begin{align}
\begin{split}
U_{ph}(T) &= 3 \int_0^{\infty} d\omega D(\omega) \frac{\hbar \omega}{e^{\frac{\hbar \omega}{k_B T}} - 1} \\
&= \frac{3L}{\pi v_s} \int_0^{\infty} d\omega \frac{\hbar \omega}{e^{\frac{\hbar \omega}{k_B T}} - 1} \\
&= \frac{3 L k_B^2 T^2}{\pi v_s \hbar} \int_0^{\infty} d\bigg(\frac{\hbar \omega}{k_B T}\bigg) \frac{\frac{\hbar \omega}{k_B T}}{e^{\frac{\hbar \omega}{k_B T}} - 1} \\
&= \frac{3 L k_B^2 T^2}{\pi v_s \hbar} \bigg(\frac{\pi^2}{6}\bigg).
\end{split}
\end{align}
While the total phonon energy scales quadratically with temperature $T$, the corresponding heat capacity varies linearly in the baseline temperature, as expected for a 1D lattice:
\begin{equation}
\label{eq: heat capacity}
C_{ph}(T) = \frac{dU_{ph}}{dT} = \frac{\pi L k_B^2}{\hbar v_s} T.
\end{equation} 
Intuitively, the scaling of the heat capacity with the lattice length $L$ corresponds to a fact that a larger lattice is more resistant to temperature change. Furthermore, the inverse variation with the speed of sound is due to the fact that a higher speed of sound leads to a lower number of occupied states (due to the sharper dispersion), causing greater occupation number increase per state for a given total energy gain, in turn leading to a greater temperature increase. Similarly, the variation with the baseline temperature $T$ can be explained by the fact that a higher baseline temperature lifts the maximum occupied energy level, resulting in a higher number of occupied states and hence a lower occupation number increase per state for a given total energy gain (and thus a lower temperature increase as well). Upon absorption of a single photon of angular frequency $\omega$ by the QD-Majorana system, the temperature increase $\Delta T$ in the nanowire is thus calculated as follows:
\begin{align} \label{eq: temperature gain single photon}
\begin{split}
\hbar \omega &= \bigg(\frac{\pi L k_B^2}{\hbar v_s} T + C_s\bigg) \Delta T, \\
\Delta T &= \hbar \omega \bigg(\frac{\pi L k_B^2}{\hbar v_s} T + C_s\bigg)^{-1},
\end{split}
\end{align}
where $C_s$ is the heat capacity of the $s$-wave superconducting strips. It is therefore desirable to minimize the superconducting critical temperature when selecting the nanowire material. It is worth noting that when calculating the actual peak temperature increase, it is important to consider dissipation through the leads. We will consider the resulting attenuation of the temperature increase in detail later in this work.

\section{Energy Transfer Rate from QD-Majorana to Nanowire Bulk}
\label{sec: Energy Transfer Rate from QD-Majorana to Nanowire Bulk}

We now calculate the energy transfer rate through carrier-phonon interaction from the photoexcited carriers of the QD-Majorana system to the phonon modes of the nanowire. Fundamentally, carrier-phonon interaction requires that the Majorana mode be occupied by an electron. Consequently, the allowed QD-Majorana tensor-product transitions are $\ket{01}_+ \rightarrow \ket{01}_-$ and $\ket{11}_+ \rightarrow \ket{11}_-$ (where the $+$ and $-$ subscripts denote the higher-energy and lower-energy states, respectively). Given a carrier-phonon interaction Hamiltonian $H''$ (which will be defined later in equation \eqref{eq:Hpp}), the emission of a phonon of wavevector $q$ and branch $\mu$ requires a transition from initial state $\ket{n_i,+}$ to a final state $\ket{n_f,-}$, where $n_i \neq n_f$. This is due to the fact that in this case, the initial and final states feature the same phase between $\ket{01}$ and $\ket{11}$, resulting in constructive interference between the $\ket{01}_+ \rightarrow \ket{01}_-$ and $\ket{11}_+ \rightarrow \ket{11}_-$ phonon emission processes:
\begin{widetext}
\begin{align}
\begin{split} \label{eq: phonon emission composite matrix element}
&\Big|\braket{(n_f,-),n_{\mu,q} + 1|H''|(n_i,+),n_{\mu,q}}\Big| \\
&= \frac{1}{4} \bigg|\Big(\bra{01,n_{\mu,q} + 1} \pm \bra{11,n_{\mu,q} + 1}\Big) H'' \Big(\ket{01,n_{\mu,q}} \pm \ket{11,n_{\mu,q}}\Big)\bigg| \\
&= \frac{1}{2} \Big|\braket{-,n_{\mu,q} + 1|H''|+,n_{\mu,q}}\Big|,
\end{split}
\end{align}
\end{widetext}
where $n_{\mu,q}$ represents the phonon number in a mode at wavevector $q$ and branch $\mu$, $(n_i,n_f) = (A,B)$, $(B,A)$, $(C,D)$, or $(D,C)$, and we have introduced the notation $\ket{+}$ and $\ket{-}$ to distinguish the higher-energy and lower-energy QD states. On the other hand, $\ket{n_i,+}$ and $\ket{n_i,-}$ feature opposite phases for $\ket{01}$ and $\ket{11}$, causing the probability amplitudes for the $\ket{01}_+ \rightarrow \ket{01}_-$ and $\ket{11}_+ \rightarrow \ket{11}_-$ phonon emission processes to cancel out through destructive interference:
\begin{widetext}
\begin{align}
\begin{split}
&\Big|\braket{(n_i,-),n_{\mu,q} + 1|H''|(n_i,+),n_{\mu,q}}\Big| \\
&= \frac{1}{4} \bigg|\Big(\bra{01,n_{\mu,q} + 1} \mp \bra{11,n_{\mu,q} + 1}\Big) H'' \Big(\ket{01,n_{\mu,q}} \pm \ket{11,n_{\mu,q}}\Big)\bigg| \\
&= 0.
\end{split}
\end{align}
\end{widetext}
Consequently, just half of the possible pairs of initial higher-energy and final lower-energy states are available for phonon-induced transitions. This will attenuate the overall transition rate by a factor of 2, which we will incorporate into Fermi's Golden Rule later in this section.

Next, we estimate the wavefunction for the QD state so that we can decompose it into plane wave states with well-defined momenta and thereby calculate the matrix element $\braket{-, n_{\mu,q} + 1|H''|+, n_{\mu,q}}$ for any phonon wavevector $q$. Although we modeled this wavefunction as an exponential function in Appendix~\ref{sec: Calculating the Dipole Matrix Elements}, its narrowness relative to the length of the nanowire ensures that it can be approximated as the square-root of a Dirac delta function from the perspective of the full nanowire:
\begin{equation}
\psi_{QD}(x) =
\begin{cases}
\frac{1}{\sqrt{b}}, & 0 < x < b \\
0, & \textrm{otherwise}
\end{cases},
\end{equation}
where $b$ denotes the span of the QD. Following the treatment in our prior analysis of Cd\textsubscript{3}As\textsubscript{2} \cite{chatterjee2021microwavephoton}, this localized state decomposes into plane-wave states $\ket{k}$ delocalized along the span of the nanowire (where $k$ is the wavevector corresponding to a particular plane-wave state) with approximately equal weight for each wavevector:
\begin{equation}
\ket{QD} = \sqrt{\frac{b}{L}} \sum_{k = -\pi/b}^{\pi/b} \ket{k}.
\end{equation}
As this expression shows, a spatially narrower edge state corresponds to a wider range of momenta, and vice versa, thus satisfying the Heisenberg uncertainty principle. This representation is useful since, in general, the carrier-phonon Hamiltonian $H''$ couples initial wavevector $k$ and final wavevector $k-q$ through emission (absorption) of a phonon $q$ ($-q$) from branch $\mu$:
\begin{equation}
H'' = \sum_{\mu,k,q} \hbar g_{\mu,q} c_{k-q}^{\dag} c_{k} \Big(b_{\mu,q}^{\dag} + b_{\mu,-q}\Big), \label{eq:Hpp}
\end{equation}
where $c_{k'} = \ket{0}\bra{k'}$ is defined as the annihilation operator for the electron at wavevector $k'$. Note that the coupling coefficient $g_{\mu,q}$ is independent of the initial wavevector $k$ (see Appendix A of \cite{chatterjee2021microwavephoton} for quantitative proof). As a result, the matrix element corresponding to the transition from $\ket{+}$ to $\ket{-}$ via phonon emission can be simplified in the following manner:
\begin{align}
\begin{split}
&\braket{-, n_{\mu,q} + 1|H''|+, n_{\mu,q}} \\
&=  \frac{a}{L} \sum_{k} \braket{k-q,n_{\mu,q} + 1|H''|k,n_{\mu,q}} \\
&= \hbar \frac{a}{L} g_{\mu,q} \sqrt{n_{\mu,q} + 1} \sum_{k = -\pi/a}^{\pi/a} 1.
\end{split}
\end{align}
Since the reciprocal space between $k = -\pi/a$ and $\pi/a$ is divided into $L/a$ segments (each of length $2\pi/L$), the transition matrix element from $\ket{+}$ to $\ket{-}$ due to emission of a phonon $q$ is simply equivalent to the transition from any initial wavevector $k$ to final wavevector $k-q$ through the same process:
\begin{align} \label{eq: carrier-phonon matrix element}
\begin{split}
&\braket{-, n_{\mu,q} + 1|H''|+, n_{\mu,q}} \\
&= \hbar g_{\mu,q} \sqrt{n_{\mu,q} + 1} \\
&= \braket{k-q,n_{\mu,q} + 1|H''|k,n_{\mu,q}}.
\end{split}
\end{align}
Next, we examine the values of the coefficients $g_{\mu,q}$. Here, we note that since the energy gap between $\ket{+}$ and $\ket{-}$ corresponds to the long-wavelength acoustic phonon regime, we can use the deformation potential treatment \cite{bardeen1950deformationpotentials}:
\begin{equation} \label{eq: deformation potential}
\sum_{\mu} |\hbar g_{\mu,q}|^2 = \frac{\hbar D^2}{2 \rho v_s V} |q|,
\end{equation}
where $D$, $v_s$ and $V$ represent the nanowire deformation potential, speed of sound, and quantization volume, respectively. 

We are now ready to determine the nonradiative decay rate. In general, for a continuum of electronic states, a rapid carrier-carrier rethermalization occurs first, elevating the system to a hot electron Fermi-Dirac distribution \cite{mihnev2016microscopicorigins, sun2008ultrafastrelaxation, dawlatya2008measurementultrafast, lundgren2015electroniccooling}. This is then followed by heat transfer from the hot electron distribution to the phonon modes via electron-phonon interaction, bringing the electron and phonon temperatures to equilibrium. Here, however, we have a discrete two-level electronic spectrum. Therefore, the sole non-negligible means of nonradiative decay is through phonon emission, which disturbs the Bose-Einstein phonon distribution and immediately gives rise to a rethermalization of the phonon modes. To this end, the phonon emission rate by an electron in $\ket{+}$ is calculated through Fermi's Golden Rule, based on the composite matrix element calculated in Eq.~\eqref{eq: phonon emission composite matrix element}:
\begin{widetext}
\begin{align}
\begin{split}
\Gamma_\mathrm{nr} &= \frac{2\pi}{\hbar} \sum_{\mu,q} \frac{1}{2} \Bigg(\frac{1}{2} \Big|\braket{-, n_{\mu,q} + 1|H''|+, n_{\mu,q}}\Big|\Bigg)^2 \delta\Big(E_+ - E_- - \hbar v_s |q|\Big) \\
&= \frac{\pi}{4\hbar} \sum_{\mu} \Big|\braket{-, n_{\mu,\pm \omega/v_s} + 1|H''|+, n_{\mu,\pm \omega/v_s}}\Big|^2 \frac{D(\omega)}{\hbar},
\end{split}
\end{align}
\end{widetext}
where $\omega = (E_+ - E_-)/\hbar$. Note that the extra factor of $1/2$ in the first line is inserted due to the fact that for an excited photoelectron in a given higher-energy state, there is only a 50\% probability that the required final state is unoccupied. The Dirac delta function enforces energy conservation, ensuring that the emitted phonon carries an frequency of $\omega = v_s q$, where $q$ is the phonon wavevector. We thus replaced the summation of the Dirac delta functions with the density of phonon modes with respect to energy, i.e. $D(\omega)/\hbar$, at $q = \pm \omega/v_s$. Using the matrix element value calculated in Eqs.~\eqref{eq: carrier-phonon matrix element} and~\eqref{eq: deformation potential}, and substituting the nanowire phonon density of states from Eq.~\eqref{eq: phonon density of states}, we find the following expression for the nonradiative decay rate at baseline nanowire temperature $T$:
\begin{align} \label{eq: nonradiative decay rate}
\begin{split}
\Gamma_\mathrm{nr}(\omega,T) &= \frac{\pi}{4\hbar} \bigg(\frac{\hbar D^2}{2 \rho v_s V} \frac{\omega}{v_s}\bigg) \Big(n(\omega,T) + 1\Big) \bigg(\frac{L}{\pi \hbar v_s}\bigg) \\
&= \frac{D^2 \omega}{8 \hbar \rho A v_s^3} \Big(\Big(e^{\frac{\hbar \omega}{k_B T}} - 1\Big)^{-1} + 1\Big),
\end{split}
\end{align}
where $n(\omega,T)$ is the Bose-Einstein phonon occupation number at frequency $\omega$ and temperature $T$, and $A$ denotes the cross-sectional area of the nanowire. We note that the nonradiative decay rate rises with increasing temperature, as expected due to the greater phonon occupation number at higher temperatures. Furthermore, in the limit $e^{\hbar \omega/k_B T} \gg 1$, it scales linearly with the phonon frequency $\omega$, since the transition strength (i.e. the matrix element amplitude-squared) increases linearly with the wavevector magnitude $|q|$. Finally, the rate varies inversely with the nanowire cross-sectional area $A$. This can be conceptualized as follows: A broader area leads to a reduction in the vibrational amplitude of each bond for a given mode energy (thus lowering the electron-phonon coupling strength per phonon), without simultaneously increasing the number of phonon modes (in the 1D limit).

\section{Ensuring Deterministic Photon Number Detection}
\label{sec: Ensuring Deterministic Photon Number Detection}

Here, we seek to verify the deterministic nature of the measurement process in two steps: first, by ensuring that the absorbed photon energy is faithfully transferred to the internal energy of the nanowire, and second, by designing a network of QD-nanowire complexes on-chip inside a cavity such that multiple absorbers can be effectuated in parallel, and the system will have a vast number of opportunities to absorb each photon (since each photon travels back and forth inside the cavity). For the former step, it is essential to calculate the parasitic radiative loss rate in order to ensure that it is negligible compared to the energy transfer rate from the photoelectrons to the bulk phonons (calculated in the previous section). For the latter step, we will derive the absorption rate per QD-nanowire complex, from which we can determine the overall probability that a photon inside the cavity is absorbed by the system before it escapes the cavity. 

Both the radiative loss and absorption rates vary quadratically with the transition dipole moment amplitude $|d_{+,-}|$ (which was calculated in Appendix~\ref{sec: Calculating the Dipole Matrix Elements}). The radiative decay rate is determined as follows \cite{hilborn1982einsteincoefficients}:
\begin{equation}
\Gamma_\mathrm{rad} = \frac{\omega^3}{3\pi \epsilon_0 \hbar c^3} |d_{+,-}|^2.
\end{equation}
Note that the radiative loss rate scales cubically with the resonance frequency. Since the microwave frequency range we are interested in falls about 5 orders of magnitude below the typical optical frequency, we would expect the rate to be far smaller than the radiative loss rate for an optical transition. In the next section, we will numerically demonstrate that the radiative loss rate is multiple orders of magnitude smaller than the phonon emission rate, given practical material parameters.

Next, we seek to derive the absorption rate for a QD-nanowire complex and lay out a procedure for calculating the total absorption probability for a system of QD-nanowires on-chip inside a cavity. For a single complex, the photon absorption rate $\Gamma_\mathrm{abs}$ is determined using Fermi's Golden Rule:
\begin{widetext}
\begin{align}
\begin{split}
&\Gamma_\mathrm{abs} \\
&= \Big(f_-(\omega,T) - f_+(\omega,T)\Big) \frac{2\pi}{\hbar} \sum_{n_i,n_f} \Big|\braket{(n_f,+),n-1|H'|(n_i,-),n}\Big|^2 \delta(E_+ - E_- - \hbar \omega) \\
&= \Big(f_-(\omega,T) - f_+(\omega,T)\Big) \frac{8\pi |g|^2 n}{\Gamma_+ + \Gamma_-},
\end{split}
\end{align}
\end{widetext}
where the parameters $f_{\pm}(\omega,T)$ denote the equilibrium populations of the upper and lower levels, respectively, of the QD-Majorana hybridization ladder, and $\Gamma_{\pm}$ represent the spectral broadening of the respective levels. In a lattice featuring a continuum of electronic states, the populations would be governed by the Fermi-Dirac distribution, with the thermal broadening dominated by electron-electron interaction. However, in this highly localized QD-Majorana system, the lack of a significant electron population strongly suppresses electron-electron interaction, leaving the dominant thermal broadening mechanism as the interaction between the QD-Majorana electrons and the nanowire phonons. To this end, the equilibrium electron populations are achieved when the phonon emission rate by upper-level electrons is balanced out by the phonon absorption rate by lower-level electrons. In turn, the phonon absorption and emission rates are proportional to $n(\omega,T)$ and $n(\omega,T) + 1$, respectively, where $n(\omega,T)$ represents the phonon number for a mode featuring a frequency $\omega$ with a lattice temperature $T$. Using the Bose-Einstein distribution to model the phonon population, we find the following relationship between $f_+$ and $f_-$:
\begin{equation}
f_+(\omega,T) \Big(\Big(e^{\frac{\hbar \omega}{k_B T}} - 1\Big)^{-1} + 1\Big) = f_-(\omega,T) \Big(e^{\frac{\hbar \omega}{k_B T}} - 1\Big)^{-1}.
\end{equation}
Furthermore, in the low-temperature limit, each two-level system consists of a single ground-state electron and a vacuum excited state. Consequently, the total electron population for each two-level system should be one, with the individual-state population representing the probability that the single electron is residing in that state:
\begin{equation}
f_+(\omega,T) + f_-(\omega,T) = 1.
\end{equation}
The above 2 expressions yield the following state populations:
\begin{align}
\label{eq: f_-}
f_-(\omega,T) &= \frac{1}{1 + e^{-\frac{\hbar \omega}{k_B T}}}, \\
\label{eq: f_+}
f_+(\omega,T) &= \frac{1}{1 + e^{\frac{\hbar \omega}{k_B T}}}.
\end{align}
As expected, for the low-temperature limit ($k_B T \ll \hbar \omega$), all of the population is concentrated in the ground state (i.e, $f_- \rightarrow 1$ and $f_+ \rightarrow 0$), whereas in the high-temperature limit ($k_B T \gg \hbar \omega$), the spectral broadening makes the states indistinguishable in population (i.e., $f_-,f_+ \rightarrow 1/2$.

We now turn to the spectral broadening of the upper and lower states ($\Gamma_+$ and $\Gamma_-$, respectively). As we discussed previously, the electron-phonon interaction serves as the dominant decay mechanism for both states. Consequently, $\Gamma_+$ and $\Gamma_-$ are approximately equivalent to the phonon emission and absorption rates, respectively, at frequency $\omega$ and temperature $T$. Based on Eq.~\eqref{eq: nonradiative decay rate}, this yields the following spectral broadening values:
\begin{align}
\label{eq: Gamma_+}
\Gamma_+ &\approx \frac{D^2 \omega}{8 \hbar \rho A v_s^3} \Big(\Big(e^{\frac{\hbar \omega}{k_B T}} - 1\Big)^{-1} + 1\Big), \\
\label{eq: Gamma_-}
\Gamma_- &\approx \frac{D^2 \omega}{8 \hbar \rho A v_s^3} \Big(e^{\frac{\hbar \omega}{k_B T}} - 1\Big)^{-1}.
\end{align}
Comporting with intuition regarding thermal broadening, a higher temperature leads to a higher spectral broadening, and vice versa.

We are thus in a position to derive a closed-form expression for the single-pass absorption probability $P_\mathrm{abs} = \Gamma_\mathrm{abs} l/c$ for a single photon in a cavity of length $l$ and beam area $A_\mathrm{beam}$ (such that the effective cavity volume $V = A_\mathrm{beam} l$), substituting state populations $f_{\pm}(\omega,T)$ from Eqs.~\eqref{eq: f_-} and~\eqref{eq: f_+}, photon-material coupling coefficient $g$ from Eq.~\eqref{eq: QD-photon coupling coefficient}, and spectral broadening values $\Gamma_{\pm}$ from Eqs.~\eqref{eq: Gamma_+} and~\eqref{eq: Gamma_-}:
\begin{equation}
P_\mathrm{abs} = \bigg(\frac{\sinh{x}}{1 + \cosh{x}}\bigg) \bigg(\frac{e^x - 1}{e^x + 1}\bigg) \frac{64\pi \rho v_s^3}{\epsilon_0 D^2 c} \frac{A}{A_\mathrm{beam}} |d_{+,-}|^2,
\end{equation}
where $x = \hbar \omega/k_B T$ represents the ratio between the gap energy and the thermal energy parameter. Note that the single-pass absorption probability for each QD-nanowire complex varies inversely with the beam area $A_\mathrm{beam}$. Consequently, for an array of complexes on a 2D chip, the total single-pass absorption probability will vary with the spatial density of complexes $\sigma = N/A_\mathrm{beam}$, where $N$ is the number of complexes covered by the beam:
\begin{align} \label{eq: absorption probability}
\begin{split}
P_\mathrm{abs,chip} &= P_\mathrm{abs} N \\
&= \bigg(\frac{\sinh{x}}{1 + \cosh{x}}\bigg) \bigg(\frac{e^x - 1}{e^x + 1}\bigg) \frac{64\pi \rho A v_s^3}{\epsilon_0 D^2 c} |d_{+,-}|^2 \sigma.
\end{split}
\end{align}
The optimal method for achieving deterministic photon absorption is by placing the chip in a high-finesse cavity. In general, Bragg mirrors can feature transmittance rates as low as 1 ppm \cite{BraggMirrorsMinimumTransmittance}. Labeling this single-pass loss probability as $P_\mathrm{loss}$, we calculate the overall absorption probability as follows:
\begin{align}
\begin{split}
P_\mathrm{abs,net} &= P_\mathrm{abs,chip} \sum_{n = 0}^{\infty} \Big(1 - P_\mathrm{abs,chip} - P_\mathrm{loss}\Big)^n \\
&= \frac{1}{1 + P_\mathrm{loss}/P_\mathrm{abs,chip}}.
\end{split}
\end{align}
In the next section, we will calculate the numerical value for the net absorption probability $P_\mathrm{abs,chip}$ given a maximal on-chip complex density $\sigma$. The results will demonstrate near-deterministic photon absorption by the detector system.

\section{Optimizing Parameters}
\label{sec: Optimizing Parameters}

Here, we provide a recipe for optimizing the controllable parameters, namely the nanowire dimensions, the nanowire material, the $s$-wave superconducting material used to induce superconductivity in the nanowire (and the associated critical temperature), and the baseline temperature relative to the critical temperature. We also calculate the temperature increase and corresponding resistance increase for a single absorbed photon, accounting for thermal dissipation.

We start by discussing the trade-offs when choosing an $s$-wave superconductor based on the critical temperature. This serves as the key temperature parameter, since the baseline temperature is only about 10-20\% lower than the critical temperature (as we will discuss later in this section). The primary advantage of a lower temperature is more deterministic detection, due to a greater photon absorption rate. The increase in the photon absorption rate is due to two effects of suppressing the electron-phonon interaction rate: first, the spectral broadening of the hybridized QD-Majorana states is reduced, thus sharpening the absorption peak; second, the population contrast between the upper and lower hybridized states is increased, thus causing the raw photon absorption process to dominate more strongly over stimulated emission. On the other hand, the key advantage of a higher critical temperature is a larger superconducting gap, which allows for detection of photons in a wider range of frequencies.

Another controllable parameter is the nanowire cross-sectional area $A$, which plays an important role in the electron-phonon interaction rate. In order to suppress the spectral broadening of the QD-Majorana system (thus ensuring a high absorption rate and a deterministic detection process), we set $A$ high enough such that the nonradiative decay rate $\Gamma_\mathrm{nr}$ (see Eq.~\eqref{eq: nonradiative decay rate}) is much lower than the resonance frequency $\omega$:
\begin{align}
\begin{split}
\frac{A}{\Big(e^{\frac{\hbar \omega}{k_B T}} - 1\Big)^{-1} + 1} & \gg \frac{D^2}{8 \hbar \rho v_s^3}.
\end{split}
\end{align}
On the other hand, we set $A$ low enough such that the phonon modes behave as a true 1D system, i.e. the energy gap between phonon wavevectors separated along the transverse axis is much greater than the thermal energy $k_B T$, thereby ensuring that the phonon population in the transverse-propagating branches is negligible and the heat capacity is minimized (thus optimizing the resolution of the detector):
\begin{align}
\begin{split}
k_B T &\ll \hbar v_s \frac{2\pi}{\sqrt{A}}, \\
T \sqrt{A} &\ll \frac{h v_s}{k_B}.
\end{split}
\end{align}
Together, these two conditions set a range for the nanowire diameter (which approximately equals $\sqrt{A}$) in terms of the baseline temperature $T = r T_c$ (where $r$ is the ratio between the baseline temperature and critical temperature) and the material properties:
\begin{equation} \label{eq: sqrtA inequality}
\frac{D}{2 \sqrt{2}} \sqrt{\frac{1}{\hbar \rho v_s^3}} \Big(\Big(e^{\frac{\hbar \omega}{k_B T}} - 1\Big)^{-1} + 1\Big)^{1/2} \ll \sqrt{A} \ll \frac{h v_s}{k_B T}.
\end{equation}
Based on the example of InAs (indium arsenide) as a typical nanowire, we use the material parameters $v_s \approx 4500 \textrm{ m}/\textrm{s}$ \cite{mariager2010acousticoscillations}, $\rho \approx 5700 \textrm{ kg}/\textrm{m}^3$, and $D \approx 6.0 \textrm{ eV}$ \cite{vurgaftman2001bandparameters}. Here, it is worth noting that the high speed of sound relative to other materials (such as InSb nanowires, which feature a speed of about 2900 m/s \cite{jurgilaitis2014timeresolved}) is advantageous for the same reason as a lower baseline temperature: it provides a greater detector absorption probability (see Eq.~\eqref{eq: absorption probability}.

We now consider the optimal range of baseline temperatures, which is related to the critical temperature of the superconducting metal used to induce superconductivity in the nanowire. Here, it is essential to choose a material such that the superconducting gap (the energy required to break a Cooper pair) is significantly larger than the photon frequency in order to ensure that the photon can only be absorbed by the QD-Majorana complex at either edge rather than by the nanowire bulk. For proximity-induced superconductivity in a nanowire, studies have shown that the critical temperature is similar to the critical temperature of the metal inducing the superconductivity \cite{pendharkar2021paritypreserving}. Quantitatively, this condition corresponds to the following expression for the superconducting critical temperature $T_c$:
\begin{equation}
T_c \gg \frac{\hbar \omega}{3.5 k_B}
\end{equation}
For a resonance frequency of 5 GHz (i.e., $\omega = \pi \times 10^{10} \textrm{ s}^{-1}$), a reasonable low-end value for $T_c$ would be 0.39 K, i.e. the critical temperature for titanium \cite{matthias1963superconductivity}, corresponding to a Cooper-pair-breaking frequency of $3.5 k_B T_c/h = 28 \textrm{ GHz}$ (far above the 5-GHz photon frequency). This lies far above the 5-GHz photon frequency, allowing for a detectable photon frequency range up to about 10 GHz. The baseline temperature should be somewhat below the critical temperature, since the superconducting gap will then approach the maximum value while the rate of change of resistance over temperature is still significant. As such, we can consider the low-end value for the baseline temperature to be roughly 0.3 K. On the other hand, a reasonable high-end value for the baseline temperature is about 0.7 K, since this is roughly the highest temperature for which a large range of nanowire diameters is available (see Eq.~\eqref{eq: sqrtA inequality}). This would allow for a detectable photon frequency range up to about 20 GHz.

Before we proceed to determining the absorption probability and detection resolution as functions of temperature and nanowire diameter, we compare the nonradiative decay rate to the radiative loss rate in order to ensure that the energy transfer from excited photoelectrons to bulk phonons is far faster than parasitic spontaneous emission. We specifically calculate the minimum nonradiative decay rate, which is applicable at the zero-temperature limit for a maximally wide nanowire. Since nanowires typically feature an upper-bound diameter of about 100 nm, we use the corresponding cross-sectional area to determine the nonradiative decay rate at the zero-temperature limit from Eq.~\eqref{eq: nonradiative decay rate}:
\begin{align}
\Gamma_\mathrm{nr,min} &= \frac{D^2 \omega}{8 \hbar \rho A v_s^3} \\
&= 8.4 \times 10^6 \textrm{ s}^{-1}.
\end{align}
On the other hand, based on the dipole moment matrix element amplitude of $|d_{+,-}| = 1.8 \times 10^{-26} \textrm{ C} \cdot \textrm{m}$ (as calculated in Appendix~\ref{sec: Calculating the Dipole Matrix Elements}), the radiative decay rate is determined using the expression laid out in Sec.~\ref{sec: Ensuring Deterministic Photon Number Detection}:
\begin{align} \label{eq: radiative decay rate}
\begin{split}
\Gamma_\mathrm{rad} &= \frac{\omega^3}{3\pi \epsilon_0 \hbar c^3} |d_{+,-}|^2 \\
&= 4.1 \times 10^{-2} \textrm{ s}^{-1},
\end{split}
\end{align}
for $\omega = \pi \times 10^{10} \textrm{ s}^{-1}$. The radiative loss rate is thus negligible even compared to the minimum nonradiative decay rate. As a result, the excited photoelectron in the QD-Majorana system will decay through the desired phonon emission channel rather than through the undesired spontaneous photon emission process.

\begin{figure}[!tb]
	\centering
	\includegraphics[width=\linewidth]{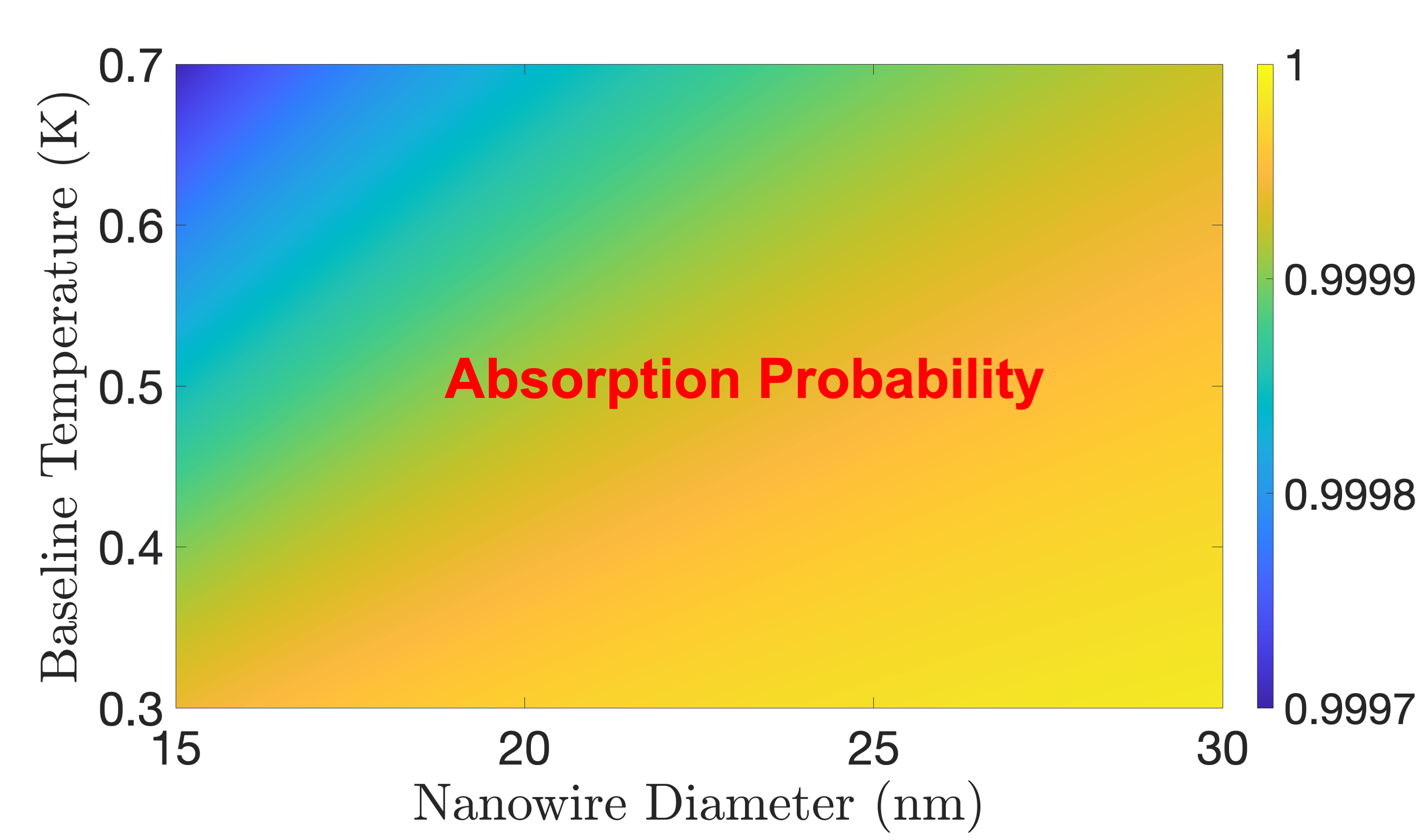}
	\caption{Photon absorption probability vs. baseline temperature and nanowire diameter for a chip consisting of InAs-QD systems inside a high-finesse cavity. Note that we assume transverse and longitudinal spacing values of 2 and 20 $\mu$m (respectively) between nanowires and a resonance frequency of 5 GHz.}
	\label{fig:absorptionprob}
\end{figure}

Having established that the energy transfer from the excited photoelectrons to the nanowire phonons is dominant, we now return to our analysis of the absorption probability as a function of temperature and nanowire diameter in order to prove that the overall detection process is deterministic. As discussed previously, we select the baseline temperature range 0.3 to 0.7 K. For the diameters, we select the values roughly satisfying Eq.~\eqref{eq: sqrtA inequality} for all temperatures up to 0.7 K. This yields a range of approximately 15 to 30 nm. In designing the chip consisting of QD-nanowire complexes, we maximize the spatial density of nanowires by setting the transverse and longitudinal unit cell dimensions to physically feasible minimum values of 2 and 20 $\mu$m, respectively. This ensures that the spacing between nearest-neighbor nanowires is far greater than the coherence length of the Majorana wavefunction (discussed later in this section). Note also that there exist 2 QD-Majorana hybridized systems per nanowire, since each nanowire features quantum dots on both ends (see Fig.~\ref{fig:qdnanowirediagram}). These parameters yield a maximum achievable QD-Majorana on-chip density of $5 \times 10^{10} \textrm{ m}^{-2}$. Figure~\ref{fig:absorptionprob} thus depicts the absorption probability (i.e., the detector efficiency) for the given temperature and diameter ranges.
As desired, the efficiency significantly exceeds 99.9\% for all design parameters shown in Figure \ref{fig:absorptionprob}, indicating an extremely deterministic detector. Also, as expected, the efficiency increases with nanowire diameter while decreasing with baseline temperature, since higher diameter and lower temperature values lower the nonradiative decay rate, thus yielding sharper absorption peaks. Recall that a lower temperature also creates a greater contrast in equilibrium population between upper and lower hybridized states, thus increasing the raw absorption rate relative to the stimulated emission rate.

Next, we consider the resolution of the detector by calculating the temperature increase per absorbed photon. Here, it is important to consider the role played by thermal dissipation, which will serve to lower the peak temperature increase. As depicted in Fig.~\ref{fig:systemdiagram}, each nanowire is surrounded by vacuum and a thermally insulating buffer. However, a pair of superconducting leads (typically made from aluminum) is continuously connected to the nanowire in order to dynamically measure the longitudinal resistance, with the other end of each lead connecting to a thermal reservoir. These leads thus serve as the dominant channel of thermal dissipation. Given a length $l_\mathrm{lead}$, cross-sectional area $A_\mathrm{lead}$, and thermal conductivity $k_t$ for each lead, the rate constant for thermal dissipation through the 2 leads can be calculated in terms of the nanowire heat capacity $C_{ph}(T)$ and the heat capacity $C_s$ of the $s$-wave superconducting strips as follows:
\begin{equation}
\label{eq: dissipation rate}
\gamma_\mathrm{dis} = \frac{2 k_t A_\mathrm{lead}}{(C_{ph}(T) + C_s) l_\mathrm{lead}},
\end{equation}
where we substituted the relationship between $C_{ph}$ and the temperature $T$ shown in Eq.~\eqref{eq: heat capacity}. The $s$-wave heat capacity $C_s$ can be calculated to first order for a generic metal using the expression determined by Phillips for aluminum \cite{phillips1959heatcapacity}, since the Fermi temperatures for the relevant metals are similar to first order:
\begin{align}
\begin{split}
C_s \approx 7.1 \gamma T_c n V_s e^{-1.34 T_c/T},
\end{split}
\end{align}
where $n$ denotes the molar density of the material (i.e. $10^5$ mol/$\textrm{m}^3$ for aluminum, corresponding to 4 atoms per unit cell), $V_s$ denotes the volume of the material, and $\gamma = 1.35 \times 10^{-3} \textrm{ J}/(\textrm{mol} \cdot \textrm{K}^2)$. Regarding the volume, the material covers about $4/5$ of the nanowire length, with a thickness as low as $d_s = 5 \textrm{ nm}$, yielding $V_s \approx 0.8 d_s L \sqrt{A}$, where $L$ and $\sqrt{A}$ are the nanowire length and diameter, respectively. Finally, as will be discussed later in this section, the baseline temperature $T$ lies slightly below the critical temperature $T_c$, with a relationship of approximately $T \approx 0.86 T_c$. Substituting these, we find that $C_s$ is linear in the nanowire length $L$ and baseline temperature $T$, as is the case with the nanowire heat capacity $C_{ph}$:
\begin{equation}
C_s \approx \zeta \sqrt{A} L T,
\end{equation}
where $\zeta = 9.4 \times 10^{-7} \textrm{ J}/(\textrm{K}^2 \cdot \textrm{m}^2)$. Based on Eq.~\eqref{eq: temperature gain single photon}, this induces the following temperature increase per photon $(\Delta T)_\mathrm{max}$ in the absence of thermal dissipation:
\begin{equation}
(\Delta T)_\mathrm{max} = \frac{\hbar \omega}{L T} \bigg(\frac{\pi k_B^2}{\hbar v_s} + \zeta \sqrt{A}\bigg)^{-1}.
\end{equation}
Per Eq.~\eqref{eq: dissipation rate}, the dissipation rate $\gamma_\mathrm{dis}$ takes the following form:
\begin{equation} \label{eq: dissipation rate 2}
\gamma_\mathrm{dis} = \frac{2 k_t A_\mathrm{lead}}{\Big(\frac{\pi k_B^2}{\hbar v_s} + \zeta \sqrt{A}\Big) l_\mathrm{lead} L T}.
\end{equation}
Given an input heat transfer rate from the absorbed photoelectron at a rate $\Gamma_\mathrm{nr}$, we can evaluate the actual temperature increase per absorbed photon as a function of ideal increase $(\Delta T)_\mathrm{max}$ using the following differential equation:
\begin{equation}
\frac{d(\Delta T)}{dt} = (\Delta T)_\mathrm{max} \Gamma_\mathrm{nr} e^{-\Gamma_\mathrm{nr} t} - \gamma_\mathrm{dis} \Delta T.
\end{equation}
Solving this, we find that the attenuation factor (i.e. the ratio between the peak temperature increase and the dissipationless ideal increase) becomes a function of $r = \gamma_\mathrm{dis}/\Gamma_\mathrm{nr}$:
\begin{equation}
\Delta T = (\Delta T)_\mathrm{max} r^{\frac{r}{1 - r}}.
\end{equation}
Note that if the dissipation rate is much higher than the nonradiative decay rate (i.e. if $r \gg 1$), then this relationship reduces to $\Delta T \approx (\Delta T)_\mathrm{max}/r$. We can obtain an analytical approximate expression for $\Delta T$ based on the dissipation rate $\gamma_\mathrm{dis}$ from Eq.~\eqref{eq: dissipation rate 2} and the nonradiative electron-phonon interaction rate $\Gamma_\mathrm{nr}$ from Eq.~\eqref{eq: nonradiative decay rate} in this regime:
\begin{equation}
\label{eq: Delta T high dissipation rate}
\Delta T \approx \frac{D^2 \omega^2 l_\mathrm{lead}}{8 \rho v_s^3 k_t A_\mathrm{lead} A} \Big(\Big(e^{\frac{\hbar \omega}{k_B T}} - 1\Big)^{-1} + 1\Big).
\end{equation}
Note that the temperature increase for the high-dissipation regime is independent of nanowire length $L$, since the higher heat capacity associated with a greater length leads to a reduced ideal temperature increased but also a lower dissipation rate, with these two shifts cancelling out. For the other two controllable parameters, i.e. nanowire cross-sectional area $A$ and baseline temperature $T$, $\Delta T$ increases with temperature and decreases with cross-sectional area. These correlations are due to the fact that the a lower cross-sectional area and a higher temperature yield a greater electron-phonon interaction rate, as previously mentioned, thus causing more rapid heat transfer into the nanowire phonons and enhancing the peak temperature increase.

\begin{figure}[!tb]
	\centering
	\includegraphics[width=\linewidth]{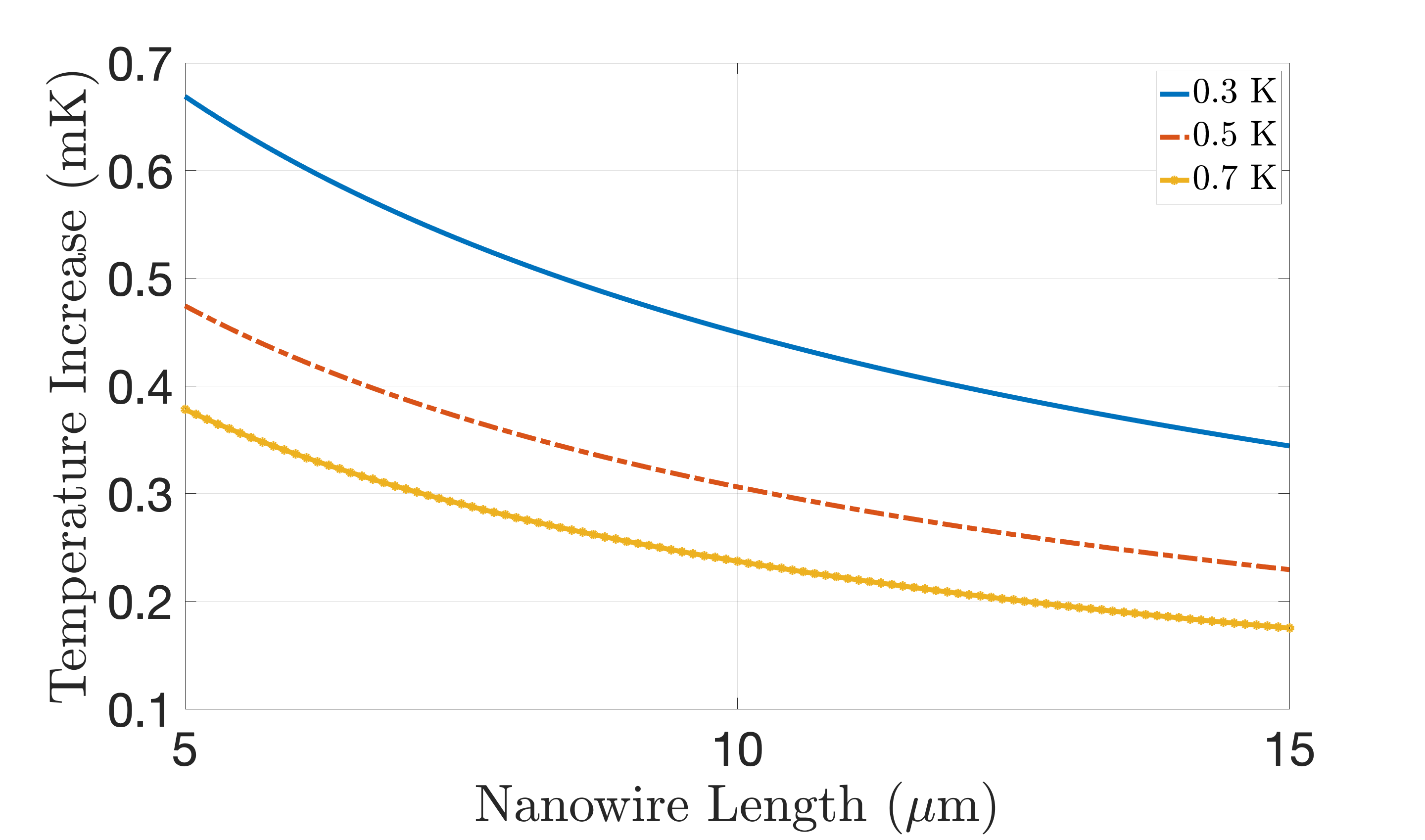}
	\caption{Single-photon temperature increase vs. nanowire length, given a 5-GHz resonance frequency and 20-nm nanowire diameter, at baseline temperatures 0.3 K (solid), 0.5 K (dot-dashed), and 0.7 K (dotted).}
	\label{fig:tempgain}
\end{figure}

\begin{figure}[!tb]
	\centering
	\includegraphics[width=\linewidth]{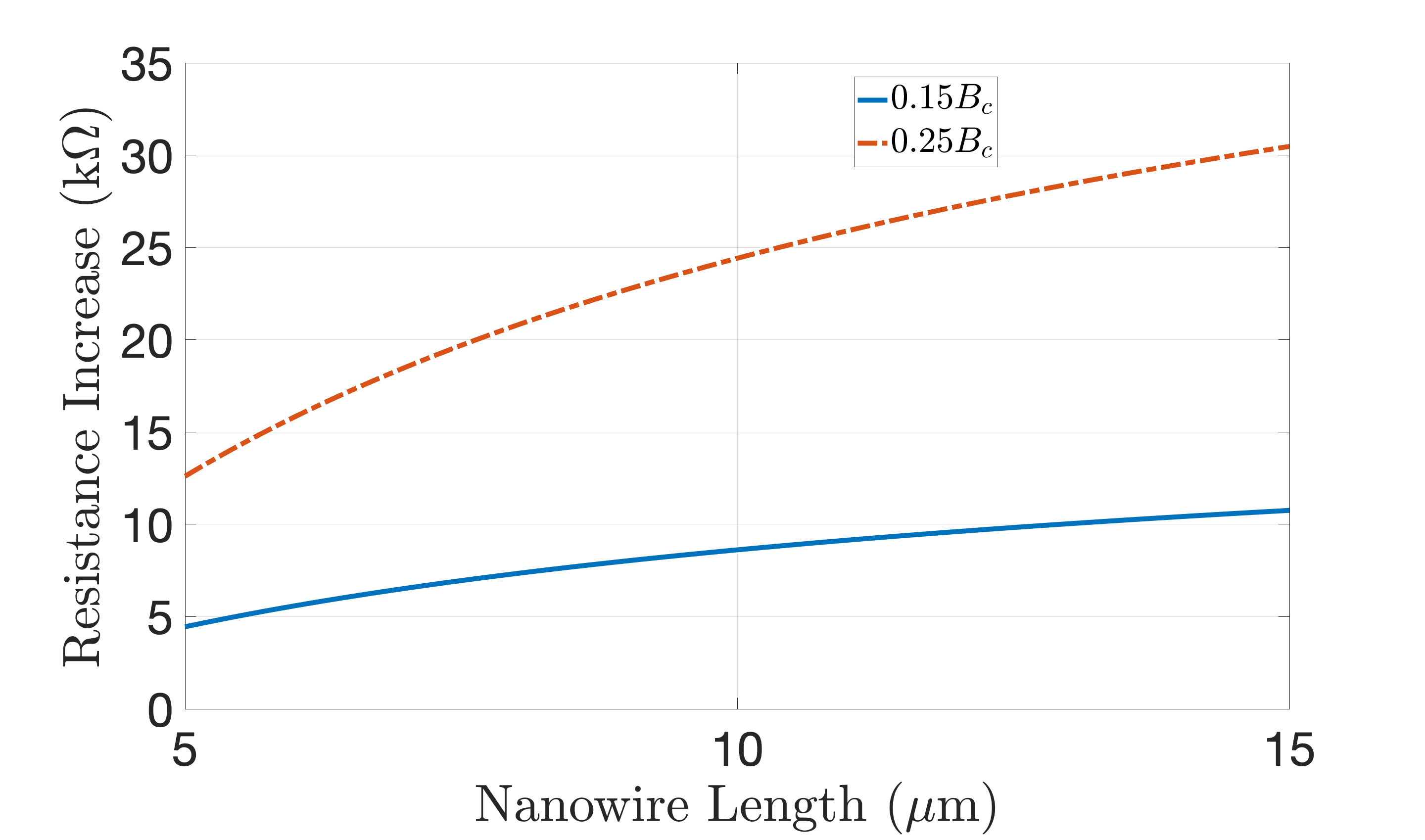}
	\caption{Single-photon resistance increase vs. nanowire length given a 20-nm nanowire diameter, 5-GHz resonance frequency, and zero-magnetic-field critical temperature $T_{c,0} = 0.56 \textrm{ K}$. The magnetic field strengths are 15\% (solid) and 25\% (dot-dashed) of the critical field $B_c$. Note that the baseline temperatures are $0.873T_{c,0}$ and $0.843T_{c,0}$, respectively, yielding $T \approx 0.5 \textrm{ K}$.}
	\label{fig:resistancegain}
\end{figure}

We now substitute practical values to numerically calculate the temperature increase per absorbed photon. Regarding the lead dimensions, each of our superconducting aluminum leads features a minimum diameter of 100 nm and a maximum length (from nanowire to thermal reservoir) of 500 $\mu$m. The thermal conductivity in the superconducting temperature range can be extrapolated from experimental data as $k_t \approx (1 \textrm{ W}/(\textrm{m} \cdot \textrm{K}^2)) T$ \cite{baudouy2014lowtemperature}. For a nanowire of diameter 20 nm, Fig.~\ref{fig:tempgain} depicts the temperature increase per photon for nanowire lengths ranging from 5 $\mu$m to 15 $\mu$m.
Note that the lower-bound value for the length is over 20 times greater than the Majorana wavefunction coherence length of 250 nm \cite{higginbotham2015quantumdots}, ensuring that the overlap between opposite Majorana states is negligible. As desired, the maximum actual temperature increase lies in the high end of the microKelvin range, thus improving the resolution far beyond the nanoKelvin range found for a cadmium arsenide (Cd\textsubscript{3}As\textsubscript{2}) detector \cite{chatterjee2021microwavephoton}. The corresponding increase in longitudinal resistance across the region between the leads can be determined from the characteristic $d\rho_e/dT$ of the material at $T$ (where $\rho_e$ is the resistivity) and the nanowire dimensions:
\begin{align}
\begin{split}
\Delta R &= \frac{L_\mathrm{between} - 2\delta}{A} \frac{d\rho_e}{dT} \Delta T \\
&= \frac{L_\mathrm{between} - 2\delta}{A} (\Delta T)_\mathrm{max} r^{\frac{r}{1 - r}} \frac{d\rho_e}{dT},
\end{split}
\end{align}
where $L_\mathrm{between}$ is the overall length of the region between the $s$-wave superconducting strips, and the parameter $\delta$ represents the distance between each end of the intermediate region and the side of the corresponding lead facing the center of the nanowire. In general, the length of each $s$-wave strip should be greater than the material's superconducting coherence length $L'$, while the length of the intermediate region should be significantly less than $2L'$ \cite{lutchyn2010majoranafermions}. Consequently, the intermediate region between the strips must take up well below $1/2$ of the total nanowire length. A useful design would be to have the intermediate region cover the middle $1/5$ of the nanowire length, with $4/5$ of the nanowire length covered by the $s$-wave strips, leaving enough room for resistance measurement through the intermediate region. This yields a strip-to-strip distance of $L_\mathrm{between} = 0.2 L$. Further, given a lead diameter of 100 nm and a spacing of 150 nm between the $s$-wave material and the side of the lead facing the material, $\delta$ reduces to 250 nm. In general, the rate of increase of resistivity with temperature, $d\rho_e/dT$, varies with the magnetic field along the nanowire. Generalizing from the results in Fig. 5(a) of Yoshizawa \textit{et al.} \cite{yoshizawa2021atomiclayer}, and setting the baseline temperature at $T = 0.9T_c$ (in order to establish a sufficient superconducting gap while also ensuring a sufficient increase of resistance with temperature) we deduce that for a magnetic field of $0.15B_c$, $d\rho_e/dT \approx 0.6\rho_\mathrm{max}/T_{c,0}$, while for a field of $0.25B_c$, $d\rho_e/dT \approx 1.7\rho_\mathrm{max}/T_{c,0}$, where $T_{c,0}$ is the critical temperature at zero magnetic field. Note that $B_c$ and $\rho_\mathrm{max}$ denote the critical magnetic field and normal-state resistivity, respectively. For the InAs nanowire, the normal-state resistivity is about $\rho_\mathrm{max} = 5.5 \times 10^{-3} \textrm{ } \Omega \cdot \textrm{m}$ \cite{zeng2018electricaltransport}, while the critical magnetic field will be set by the $s$-wave superconducting material in proximity to the InAs nanowire. Consequently, the resistance increase per photon takes the following form:
\begin{align}
\begin{split}
\Delta R &= \frac{L_\mathrm{between} - 2\delta}{A} \frac{n \rho_\mathrm{max}}{T_{c,0}} (\Delta T)_\mathrm{max} r^{\frac{r}{1 - r}},
\end{split}
\end{align}
where $n = 0.6$ and $1.7$ for $B = 0.15B_c$ and $0.25B_c$, respectively. Figure~\ref{fig:resistancegain} depicts the single-photon resistance increase as a function of nanowire length for a diameter of 20 nm and a zero-field critical temperature $T_{c,0} = 0.56 \textrm{ K}$. 
For the purpose of setting a precise baseline temperature, it is important to note that the critical temperature decreases slightly as the magnetic field is increased to $0.25B_c$, reaching values of $0.97T_{c,0}$ and $0.937T_{c,0}$ for field strengths of $0.15B_c$ and $0.25B_c$, respectively. As previously discussed, we set the baseline temperature 10\% below the critical temperature, yielding baseline values of $0.873T_{c,0}$ and $0.843T_{c,0}$ for the respective fields, corresponding to $T \approx 0.5 \textrm{ K}$. Throughout the relevant range of nanowire lengths, we find a resistance increase above in the kiloohm range, far above the minimum resolution for resistance measurement. In fact, the resistance increase per photon is on the order of 9 orders of magnitude higher than the corresponding resistance increase in a Cd\textsubscript{3}As\textsubscript{2} device \cite{chatterjee2021microwavephoton}. As such, we could set the baseline temperature even significantly below 90\% of the critical temperature. That way, we would further increase the superconducting gap (allowing for the detection of a greater range of photon frequencies) while still obtaining an easily resolvable single-photon resistance increase. 

We now briefly discuss how to tune the QD-Majorana coupling coefficient $\lambda$ in order to satisfy the resonance condition $\omega = 2\lambda$ and physically implement the scheme discussed in Appendix~\ref{sec: Calculating the Dipole Matrix Elements}. To this end, the ability to open a microwave-frequency gap in the Majorana zero mode has been recently demonstrated experimentally \cite{vanzanten2020photonassisted}. In general, $\lambda$ can be controlled by properly setting the distance between the QD electronic mode and the nanowire edge mode. It is desirable to fabricate the system such that the length of the QD yields the center-to-center distance required for the hybridization gap to be resonant with the photon frequency. However, small adjustments may be required post-fabrication in order to satisfy the resonance condition more precisely. One potential means of achieving this is by using a piezoelectric material. Specifically, such a structure can be built by fabricating the quantum dot in a pillar protruding out from the buffer material, as has been demonstrated by Oulton \cite{oulton2014electrifyingcavities}. Similarly, the nanowire can also levitated by a couple of insulating supports above the buffer. The piezoelectric material can be then be sandwiched between the base of the quantum dot pillar and the nearest support for the nanowire. By applying a small voltage, the piezoelectric material can be stretched or compressed, thus providing for the ability to tune the QD-nanowire distance. Another means of tuning the QD-Majorana coupling is by adjusting the height of the potential energy barrier separating the two wells. Appendix~\ref{sec: Fine-Tuning the Dipole Matrix Element} provides an analysis of the shift in barrier height required to induce a particular fractional change in the QD-Majorana coupling strength (and hence in the resonance frequency).

\section{Conclusion}

We have theoretically demonstrated a revolutionary ultra-high-precision microwave photon number detector using a 1D system consisting of a nanowire coupled to a quantum dot at each end. To the best of our knowledge, our system serves as the first real-world application of Majorana zero modes since topological computing. The Majorana edge states couple with the quantum dots, giving rise to a hybridized energy spectrum well-suited for microwave photon absorption, while the superconducting nanowire bulk acts as a bolometer. As with our previous Cd\textsubscript{3}As\textsubscript{2} detector \cite{chatterjee2021microwavephoton}, this system offers the benefits of complete spatial separation between the absorber and the bolometer, as well as rapid and deterministic energy transfer from the absorber electrons to the bolometer phonons. However, owing to the vastly reduced heat capacity caused by the low dimensionality, our nanowire system improves upon the Cd\textsubscript{3}As\textsubscript{2} detector's measurement precision (resistance increase per photon) by 9 orders of magnitude, serving as a major breakthrough in microwave-photon detection resolution.

In addition to serving as a photon number detector, the nanowire-QD system will provide a highly promising platform for conclusively proving the existence of Majorana zero modes. Specifically, demonstrating photon absorption would prove the existence of a zero-energy edge mode that hybridizes with the quantum dot. Furthermore, we can verify the spinlessness and charge-neutrality of such a mode by applying new magnetic and DC electric fields to test the Zeeman and Stark effects, respectively. The energy level of a spinless and charge-neutral Majorana edge mode would be unperturbed by these fields, while the quantum dot level would shift, thus breaking the QD-Majorana hybridization and suppressing photon absorption. Our nanowire-QD system thus holds significant potential in solving one of the central physics questions in recent decades.

\begin{acknowledgements}

Sandia National Laboratories is a multimission laboratory managed and operated by National Technology \& Engineering Solutions of Sandia, LLC, a wholly owned subsidiary of Honeywell International Inc., for the U.S. Department of Energy’s National Nuclear Security Ad- ministration under contract DE-NA-0003525.
	
\end{acknowledgements}

\begin{appendix}

\section{Calculating the Dipole Matrix Elements}
\label{sec: Calculating the Dipole Matrix Elements}

Here, we will calculate the dipole matrix element corresponding to transitions between lower-energy and higher-energy QD-Majorana hybrid states. It is important to note that an electromagnetic field cannot induce a transition between the vacuum state and the occupied state in either the QD or the nanowire edge alone. Instead, a field acts on the complex by shifting the electron position in a manner that induces hopping between the QD and the nanowire edge. As such, net charge must be conserved in any photon-induced transition. Consequently, the dipole moment can only act between $\ket{01}$ and $\ket{10}$, where the first and second indices represent the electron occupation numbers in the QD and Majorana modes, respectively. We thus solve the dipole matrix element for each pair by decomposing the composite states from Eqs.~\eqref{eq: A,+ state} through~\eqref{eq: B,- state} into the occupation number basis, namely $\ket{00}$, $\ket{01}$, $\ket{10}$, and $\ket{11}$, starting with the QD-Majorana pair on the left end of the nanowire:
\begin{align}
\begin{split} \label{eq: state pair A decomposition}
&\ket{A,\pm} \\
&= \frac{1}{\sqrt{2}} \bigg(\frac{1}{\sqrt{2}} \Big(\ket{00} - \ket{01}\Big) \mp \frac{1}{\sqrt{2}} \Big(\ket{10} - \ket{11}\Big)\bigg) \\
&= \frac{1}{2} \Big(\ket{00} - \ket{01} \mp \ket{10} \pm \ket{11}\Big),
\end{split}
\\
\begin{split} \label{eq: state pair B decomposition}
&\ket{B,\pm} \\
&= \frac{1}{\sqrt{2}} \bigg(\frac{1}{\sqrt{2}} \Big(\ket{00} + \ket{01}\Big) \pm \frac{1}{\sqrt{2}} \Big(\ket{10} + \ket{11}\Big)\bigg) \\
&= \frac{1}{2} \Big(\ket{00} + \ket{01} \pm \ket{10} \pm \ket{11}\Big).
\end{split}
\end{align}
In solving for the dipole matrix elements, it is conventional to express the dipole operator $\bm{d}$ in terms of the momentum operator $\bm{p}$ instead:
\begin{equation}
\braket{n_f,+|\bm{d}|n_i,-} = -\frac{i q_e}{m\omega} \braket{n_f,+|\bm{p}|n_i,-},
\end{equation}
where $q_e$ and $m$ represent the electron charge and effective mass, respectively, and $\omega = (E_+ - E_-)/\hbar$.
In terms of the hybrid states, Eqs.~\eqref{eq: state pair A decomposition} and~\eqref{eq: state pair B decomposition} imply that all transitions between higher-energy and lower-energy states, i.e. $\ket{A,-} \leftrightarrow \ket{A,+}$, $\ket{B,-} \leftrightarrow \ket{B,+}$, $\ket{A,-} \leftrightarrow \ket{B,+}$, and $\ket{B,-} \leftrightarrow \ket{A,+}$ are valid. Noting that all hopping fundamentally takes place between $\ket{01}$ and $\ket{10}$ (representing a carrier going back-and-forth between the QD and Majorana modes), and utilizing the fact that for the higher-energy (lower-energy) states, the coefficients for $\ket{01}$ and $\ket{10}$ feature equal (opposite) signs, we find that all of the transitions for the QD-Majorana complex on the left end of the nanowire feature the same matrix-element amplitude:
\begin{align} \label{eq: dipole matrix element amplitude}
\begin{split}
&\Big|\braket{n_f,+|\bm{d}|n_i,-}\Big| \\
&= \bigg|-\frac{i q_e}{m\omega} \braket{n_f,+|\bm{p}|n_i,-}\bigg| \\
&= \frac{q_e}{4 m\omega} \bigg|\Big(\bra{01} + \bra{10}\Big) \bm{p} \Big(\ket{01} - \ket{10}\Big)\bigg| \\
&= \frac{q_e}{2 m\omega} \bigg|\textrm{Im}\Big[\braket{01|\bm{p}|10}\Big]\bigg|,
\end{split}
\end{align}
where $n_i$ and $n_f$ can each equal either $A$ or $B$. Note that the result in the last line is a consequence of the Hermitian nature of the momentum operator $\bm{p}$. 

Next, we examine the dipole matrix elements for the right end of the nanowire. As in the left end, we substitute the representation of the composite states shown in Eqs.~\eqref{eq: A,+ state} through~\eqref{eq: B,- state}:
\begin{align}
\begin{split} \label{eq: state pair C decomposition}
&\ket{C,\pm} \\
&= \frac{1}{\sqrt{2}} \bigg(\frac{1}{\sqrt{2}} \Big(\ket{00} - \ket{01}\Big) \mp \frac{1}{\sqrt{2}} \Big(\ket{10} + \ket{11}\Big)\bigg) \\
&= \frac{1}{2} \Big(\ket{00} - \ket{01} \mp \ket{10} \mp \ket{11}\Big),
\end{split}
\\
\begin{split} \label{eq: state pair D decomposition}
&\ket{D,\pm} \\
&= \frac{1}{\sqrt{2}} \bigg(\frac{1}{\sqrt{2}} \Big(\ket{00} + \ket{01}\Big) \pm \frac{1}{\sqrt{2}} \Big(\ket{10} - \ket{11}\Big)\bigg) \\
&= \frac{1}{2} \Big(\ket{00} + \ket{01} \pm \ket{10} \mp \ket{11}\Big).
\end{split}
\end{align}
Note that the sign corresponding to $\ket{01}$ and $\ket{10}$ each is the same for the $A$ and $C$ superpositions, as well as for the $B$ and $D$ superpositions. Consequently, the matrix-element amplitude for each transition is equivalent to that calculated in Eq.~\eqref{eq: dipole matrix element amplitude}:
\begin{equation}
\Big|\braket{n_f,+|\bm{d}|n_i,-}\Big| = \frac{q_e}{2 m\omega} \bigg|\textrm{Im}\Big[\braket{01|\bm{p}|10}\Big]\bigg|,
\end{equation}
where $n_i$ and $n_f$ can each equal either $C$ or $D$. 

It is worth noting that these matrix elements are non-zero if and only if the QD-Majorana hopping dipole moment contains an imaginary component. Indeed, since $\bm{p} = -i\hbar \nabla$, and since $\ket{10}$ and $\ket{01}$ are both localized states with fully real wavefunction values in position space, the matrix elements are fully imaginary, as desired. Defining $\bm{\hat{y}}$ as the direction of the electric field $\bm{E}$, the matrix-element amplitudes for $\bm{d} \cdot \bm{\hat{E}}$ thus reduce to the following:
\begin{align}
\begin{split}
|d_{+,-}| &=\frac{q_e}{2 m\omega} \bigg|\textrm{Im}\Big[\braket{01|\bm{p} \cdot \bm{\hat{E}}|10}\Big]\bigg| \\
&= \frac{q_e \hbar}{2 m \omega} \Big|\braket{01|\partial_y|10}\Big|.
\end{split}
\end{align}
Although the localized states $\ket{10}$ and $\ket{01}$ effectively reduce to Dirac delta functions from the point of view of the nanowire length, it is convenient to model them as ground-state solutions to finite potential wells, with exponentially decaying tails outside their respective wells:
\begin{equation}
\psi(y) = N
\begin{cases}
\cos{(ky)}, & |y| < \frac{w}{2} \\
A e^{-\kappa |y|}, & |y| > \frac{w}{2}
\end{cases},
\end{equation}
where $\kappa$, $A$, and the normalization coefficient $N$ are functions of the well width $w$ and the constant $k$ (proportional to the square of the state energy):
\begin{align}
\kappa &= k \tan{\bigg(\frac{kw}{2}\bigg)}, \\
A &= e^{\kappa w/2} \cos{\bigg(\frac{kw}{2}\bigg)}, \\
N &= \bigg(\frac{kw + \sin{(kw)}}{2k} + \frac{1 + \cos{(kw)}}{2\kappa}\bigg)^{-1/2}.
\end{align}
The potential well depth can be expressed in terms of $k$ and $\kappa$ as $V_0 = f(k^2 + \kappa^2)$, where $f = \hbar^2/2m$. 

In order for the exponentially decaying part of each wavefunction to be dominant, it is necessary that the well width be much smaller than the wavefunction span, i.e. $w \ll 1/\kappa$. Furthermore, for the ground state, it is necessary that $w \ll 1/k$. Applying these limits, we find that the normalization coefficient $N$ approximately reduces to $\sqrt{\kappa}$, and $A \approx 1$. Then, if the QD and Majorana mode centers are separated by a length $l > 1/\kappa \gg w$, the dipole matrix element is solved by the following integral:
\begin{align} \label{eq: d_+,-}
\begin{split}
|d_{+,-}| &= \frac{q_e \hbar}{2 m \omega} \bigg|\int dy \Big(N A e^{\kappa (y-l)}\Big) \partial_y \Big(N A e^{-\kappa y}\Big)\bigg| \\
&\approx \frac{q_e \hbar N^2 A^2 \kappa}{2 m \omega} e^{-\kappa l} \bigg|\int_{w/2}^{l - w/2} dy\bigg| \\
&\approx  \frac{q_e \hbar \kappa^2 l}{2 m \omega} e^{-\kappa l},
\end{split}
\end{align}
Note that for $|y - l/2| > (l-w)/2$, the initial and final wavefunctions are both even about $y = l/2$, thus negating the contribution of these regions to the dipole matrix element. It is also worth noting that at small center-to-center distances $l$, the dipole matrix element increases with the distance, whereas at long distances, the correlation is reversed. This is due to the trade-off between an increase of dipole moment with distance between charges and the drop-off in wavefunction overlap with distance.

Next, we seek to solve for the center-to-center distance $l$ that produces the desired QD-Majorana coupling $\lambda = \omega/2$, where $\omega$ is the RF photon angular frequency. Note that $\hbar \lambda$ represents the transition amplitude for a charge carrier between the QD mode and the Majorana mode. To calculate this hopping parameter, we use as our operator the alteration of the potential energy landscape induced by the new neighboring potential well:
\begin{equation}
\Delta V(y) = 
\begin{cases}
0, & x < l - \frac{w}{2} \\
-V, & l - \frac{w}{2} < x < l + \frac{w}{2} \\
0, & x > l + \frac{w}{2}
\end{cases}.
\end{equation}
Consequently, the hopping parameter is determined as follows:
\begin{widetext}
\begin{align} \label{eq: lambda generic}
\begin{split}
\lambda &= \frac{1}{\hbar} \Big|\braket{\psi_\textrm{MZM}|\Delta V|\psi_\textrm{QD}}\Big| \\
&= \frac{\hbar}{2m} \bigg|\int_{l-w/2}^{l+w/2} dy \Big(N \cos{(k(y-l))}\Big) (-\kappa^2 - k^2) \Big(N A e^{-\kappa y}\Big)\bigg| \\
&\approx \frac{\hbar}{2m} \bigg|\int_{l-w/2}^{l+w/2} dy \Big(\sqrt{\kappa}\Big) (-k^2) \Big(\sqrt{\kappa} e^{-\kappa l}\Big)\bigg| \\
&\approx \frac{\hbar \kappa^2}{m} e^{-\kappa l},
\end{split}
\end{align}
\end{widetext}
where we used the narrow-well approximations $\kappa w \ll 1$ and $kw \ll 1 < \kappa l$, yielding $N \approx \sqrt{\kappa}$, $A \approx 1$, and $k^2 \approx 2\kappa/w \gg \kappa^2$. It is now straightforward to determine the dipole matrix element amplitude $|d_{+,-}|$ by dividing Eq.~\eqref{eq: d_+,-} by~\eqref{eq: lambda generic} and substituting $\omega = 2\lambda$:
\begin{equation}
|d_{+,-}| \approx \frac{q_e l}{4}.
\end{equation}
As intuitively expected, the dipole moment amplitude linearly varies with the center-to-center length $l$. Regarding the value of $\kappa$, although the coherence length of the Majorana wavefunction is about 14 nm in iron-based superconducting nanowires \cite{chiu2020scalablemajorana}, the corresponding length scale for InAs nanowires was determined by Higginbotham to be about 250 nm \cite{higginbotham2015quantumdots}, consistent with InAs material parameters laid forth by Das \textit{et al.} \cite{das2012zerobias}. Using the inverse of this as $\kappa$, along with an electron effective mass of 0.020$m_0$ in InAs \cite{sladek1957effectivemasses}, we find that for a frequency of 5 GHz, $l = 440 \textrm{ nm}$, corresponding to $|d_{+,-}| \approx 1.8 \times 10^{-26} \textrm{ C} \cdot \textrm{m}$. The center-to-center distance is significantly longer than the coherence length of each wavefunction, as desired. It is worth noting that both the center-to-center distance and the dipole matrix element amplitude correlate negatively with the resonance frequency. This is due to the fact that a higher resonance frequency necessitates a stronger QD-Majorana coupling strength, which in turn requires a closer spacing between the modes, thus lowering the center-to-center distance $l$. This also reduces the dipole moment amplitude due to the aforementioned attenuation at lower center-to-center spacing.

\section{Fine-Tuning the Dipole Matrix Element}
\label{sec: Fine-Tuning the Dipole Matrix Element}

Here, we discuss the adjustments required to the height of the barrier separating the QD and Majorana wells in order to tune the QD-Majorana coupling strength. This method is advantageous in that it preserves symmetry in the potential energy landscape for the two modes. Labeling the shift in the potential energy barrier as $\delta V$, we obtain the following wavefunctions for the quantum dot and Majorana states, respectively, in the narrow-well limit:
\begin{align}
\psi_\mathrm{QD}(y) &\approx \bigg(\frac{1}{2\kappa_1} + \frac{1}{2\kappa_2}\bigg)^{-1/2} 
\begin{cases}
e^{\kappa_1 x}, & x < -\frac{w}{2} \\
1, & -\frac{w}{2} < x < \frac{w}{2} \\
e^{-\kappa_2 x}, & x > \frac{w}{2}
\end{cases}, \\
\begin{split}
\psi_\mathrm{MZM}(y) &\approx \bigg(\frac{1}{2\kappa_1} + \frac{1}{2\kappa_2}\bigg)^{-1/2} \\
&\quad\times
\begin{cases}
e^{\kappa_2 x}, & x < l - \frac{w}{2} \\
1, & l - \frac{w}{2} < x < l + \frac{w}{2} \\
e^{-\kappa_1 x}, & x > l + \frac{w}{2}
\end{cases},
\end{split}
\end{align}
where $\kappa_1$ and $\kappa_2$ are defined in the following manner:
\begin{align}
\kappa_1 &= \sqrt{\frac{2mV_1}{\hbar^2} - k^2}, \\
\kappa_2 &= \sqrt{\frac{2m(V_1 + \delta V)}{\hbar^2} - k^2} = \sqrt{\kappa_1^2 + \frac{2m \delta V}{\hbar^2}}.
\end{align}
Similar to our approximation that $k^2 \approx 2\kappa/w$ for the case of $V_1 = V_2 = V$, we can show that $k^2 \approx (\kappa_1 + \kappa_2)/w$ for the general case. If we also assume a perturbative barrier potential energy shift such that $\delta V \ll V_1$, we can approximate the relationship between the decay constants $\kappa_1$ and $\kappa_2$ and the original $\kappa$ (for $\delta V = 0$) as follows:
\begin{align}
\label{eq: kappa1 approx}
\kappa_1 &\approx \kappa - \frac{m \delta V}{2 \hbar^2 \kappa} \\
\label{eq: kappa2 approx}
\kappa_2 &\approx \kappa + \frac{m \delta V}{2 \hbar^2 \kappa}
\end{align}
We are now in a position to calculate the change in the hopping parameter induced by the barrier potential energy shift $\delta V$. To this end, we use the method that we introduced in Eq.~\eqref{eq: lambda generic}:
\begin{widetext}
\begin{align} \label{eq: lambda asymmetric}
\begin{split}
\lambda &= \frac{1}{\hbar} \Big|\braket{\psi_\textrm{MZM}|\Delta V|\psi_\textrm{QD}}\Big| \\
&\approx \frac{\hbar}{2m} \bigg|\int_{l-w/2}^{l+w/2} dy \bigg(\frac{1}{2\kappa_1} + \frac{1}{2\kappa_2}\bigg)^{-1/2} (-\kappa_2^2 - k^2) \bigg(\frac{1}{2\kappa_1} + \frac{1}{2\kappa_2}\bigg)^{-1/2} e^{-\kappa_2 y}\bigg| \\
&\approx \frac{\hbar}{2m} \bigg(\frac{1}{2\kappa_1} + \frac{1}{2\kappa_2}\bigg)^{-1} \bigg|\int_{l-w/2}^{l+w/2} dy (-k^2) e^{-\kappa_2 l}\bigg| \\
&\approx \frac{\hbar \kappa_1 \kappa_2}{m} e^{-\kappa_2 l},
\end{split}
\end{align}
\end{widetext}
Defining $\delta \kappa = m \delta V/(2 \hbar^2 \kappa)$, we can make the first-order approximation that $\kappa_1 \kappa_2 \approx \kappa^2$, leading to the conclusion that the dominant effect of the shift in the potential energy barrier occurs on the exponential term due to the replacement $e^{-\kappa l} \rightarrow e^{-(\kappa + \delta k) l}$. Consequently, given the fact that $\lambda$ is proportional to the resonance frequency $\omega$, the original resonance frequency $\omega_0$ can be shifted to a different value $\omega'$ via the following potential energy barrier shift $\delta V$:
\begin{equation} \label{eq: delta V}
\delta V \approx -\frac{2 \hbar^2 \kappa}{m l} \ln{\bigg(\frac{\omega'}{\omega_0}\bigg)} 
\end{equation}
As desired, lowering the barrier ($\delta V < 0$) leads to stronger QD-Majorana hybridization and hence a larger frequency gap, and vice versa.

The dipole moment amplitude $|d_{+,-}|$ is solved in a manner analogous to Eq.~\eqref{eq: d_+,-}, focusing on the region between the wells (which dominates the overlap):
\begin{align} \label{eq: d_+,- asymmetric}
\begin{split}
&|d_{+,-}| \\
&= \frac{q_e \hbar}{2 m \omega} \Bigg|\int dy \Bigg(\bigg(\frac{1}{2\kappa_1} + \frac{1}{2\kappa_2}\bigg)^{-1/2} e^{\kappa_2 (y-l)}\Bigg) \\
&\quad \times \partial_y \Bigg(\bigg(\frac{1}{2\kappa_1} + \frac{1}{2\kappa_2}\bigg)^{-1/2} e^{-\kappa_2 y}\Bigg)\Bigg| \\
&\approx \frac{q_e \hbar}{2 m \omega} \bigg(\frac{1}{2\kappa_1} + \frac{1}{2\kappa_2}\bigg)^{-1} \kappa_2 e^{-\kappa_2 l} \bigg|\int_{w/2}^{l - w/2} dy\bigg| \\
&\approx  \frac{q_e \hbar \kappa_1 \kappa_2^2 l}{m \omega (\kappa_1 + \kappa_2)} e^{-\kappa_2 l},
\end{split}
\end{align}
Substituting $\omega = 2\lambda$, and merging Eqs.~\eqref{eq: lambda asymmetric} and~\eqref{eq: d_+,- asymmetric}, we find $|d_{+,-}|$ in terms of the center-to-center distance $l$ and the decay constants $\kappa_1$ and $\kappa_2$:
\begin{equation}
|d_{+,-}| \approx \frac{q_e l}{2} \frac{\kappa_2}{\kappa_1 + \kappa_2}.
\end{equation}
We can express the dipole moment perturbation $\delta |d_{+,-}|$ due to the barrier potential energy shift $\delta V$ required to induce the resonance frequency shift $\omega_0 \rightarrow \omega'$ by substituting the approximate expressions from Eq.~\eqref{eq: kappa1 approx} and~\eqref{eq: kappa2 approx}, as well as the expression in Eq.~\eqref{eq: delta V} relating $\delta V$ to the ratio between new and original resonance frequencies:
\begin{align}
\begin{split}
\delta |d_{+,-}| &\approx \frac{q_e l}{2} \frac{\delta \kappa}{2\kappa} \\
&\approx \frac{q_e l}{4 \kappa} \frac{m}{2 \hbar^2 \kappa} \delta V \\
&\approx -\frac{q_e}{4 \kappa} \ln{\bigg(\frac{\omega'}{\omega_0}\bigg)}.
\end{split}
\end{align}

\end{appendix}

\end{document}